\documentclass[12pt]{article}
\usepackage[hmargin=2.5cm,vmargin=3cm]{geometry}
 \usepackage{graphicx}
 \usepackage{amsmath}
 \usepackage{amssymb}
  \usepackage{enumerate}

\begin{document} \title{Time-of-arrival correlations} \author {Charis Anastopoulos\footnote{anastop@physics.upatras.gr} and   Nina Savvidou\footnote{ksavvidou@upatras.gr}\\
 {\small Department of Physics, University of Patras, 26500 Greece} }

\maketitle

\begin{abstract}We propose that measurements of time-of-arrival correlations in multi-partite systems can sharply distinguish between different approaches to the time-of-arrival problem.  To show this, we construct a Positive-Operator-Valued measure for two distinct time-of-arrival measurements in a bipartite system, and we prove that the resulting probabilities differ strongly from ones defined in terms of probability currents. We also prove that time-of-arrival correlations are entanglement witnesses, a result suggesting the use of  temporal observables  for quantum information processing tasks. Finally, we construct the probabilities for sequential time-of-arrival measurements on a single particle. We derive the state-reduction rule for time-of-arrival measurements; it differs significantly  from the standard one, because time-of-arrival measurements are not defined at a single predetermined moment of time.
\end{abstract}

\section{Introduction} The simplest  version of the time-of-arrival problem in quantum mechanics  \cite{ML, ToAbooks} is the   following. A particle is prepared on an initial state $|\psi_0 \rangle$ that is localized around $x = 0$ and has positive mean momentum. A  detector is located  at $x = L$. What is the  probability  $P(L, t)dt$ that the particle is detected at $x = L$ at some moment between $t$ and $t+\delta t$?

 In spite of the problem's apparent simplicity, there is no consensus on the answer. The reason is that there exists no self-adjoint operator for time in quantum mechanics \cite{Pauli}; hence, we cannot obtain an unambiguous answer by employing Born's rule. Several different approaches to the problem have been developed, each following a different reasoning. All approaches lead to probability densities $P(L, t)$ that differ   from each other only at the level of small quantum fluctuations, so that they cannot easily be distinguished experimentally. Such a distinction would be highly desirable, because the time-of-arrival problem  is  only an elementary  manifestation of an important   foundational issue, namely, understanding the role of time in quantum mechanics.

The main idea of this paper is that  different theories about the time-of-arrival   could be distinguished  if they are applied to more elaborate time-of-arrival measurements. Consider,  for example,  a multi-particle system. The time of arrival of each particle is a distinct observable that is recorded by different detectors.  The  correlations between different time-of-arrival observables are in principle measurable We expect that different theories will lead to different predictions for such correlations.

 We implement this idea by extending  the construction of time-of-arrival probabilities of Ref.  \cite{AnSav12} to set-ups that involve two or more time-of-arrival measurements.   We express time-of-arrival correlations in terms of Positive-Operator-Valued Measures (POVMs). These correlations  turn out to {\em differ significantly} from ones  defined in terms of probability currents.

Our treatment of the time of arrival is based on the  Quantum Temporal Probabilities (QTP) method  \cite{AnSav12, AnSav13}. The QTP method provides an algorithm, applicable to any quantum system, that allows for the construction of quantum probabilities in which time is treated as a random variable. Besides the time-of-arrival problem \cite{AnSav12, AnSav06},  the method has also been applied for the temporal characterization of tunneling  \cite{AnSav13, AnSav08, An08}, for calculating the response and correlations of accelerated particle detectors \cite{AnSav11} and to  relativistic quantum measurements \cite{AnSav15}.
The key idea  is to distinguish between the time parameter of Schr\"odinger equation from the time of occurrence of  a measurement event \cite{Sav99, Sav10}.   The latter are physical observables: they are treated as  quasiclassical macroscopic variables associated to a detector's degrees of freedom, and the associated probabilities are   unambiguously defined \cite{GeHa}.

The probability density  with respect to the times of $n$ particle detection events is a linear functional of a  $2n$--correlation functions of the associated quantum field.  The simplest time-of-arrival measurement involves
 one particle and one detector, hence, the detection probability is a linear functional of the two-point function \cite{AnSav12}.  The measurements considered in this paper involve two detection events; either one detection for each particle in a bi-partite system, or two successive detections of   a single particle. The associated probability densities  are linear functionals of   field four-point functions.

 \medskip

Our analysis proceeds as follows.
 First, we revisit the time-of-arrival probability measure of Ref. \cite{AnSav12}. We transform the measure to the classical state space, and study the properties of the quantum fluctuations. We find  that different proposals about the time  of arrival  are  distinguished at very low momenta (or equivalently, very low temperatures): the traversed distance must be of the order of the particles' thermal de Broglie wave-length.

Second, we consider time-of-arrival measurements in bipartite particle systems. We derive the joint probability distribution for the times of  arrival $t_1$ and $t_2$ of the two particles.  We prove that the resulting probabilities   cannot, in general, be expressed in terms of  probability currents. Thus, we prove that different approaches to the time of arrival lead to experimentally distinguishable predictions for time-of-arrival correlations. Moreover, we show that these correlation define {\em entanglement witnesses}. This result suggests that quantum   temporal observables can be used for the detection of entanglement and, possibly, for  information processing tasks.

Third, we consider sequential time-of-arrival measurements on a single particle. We derive the joint probability distribution for the times of arrival $t_1$ and $t_2$ at two spatially separated detectors. We identify the rule for the change of the quantum state after a time-of-arrival measurement and show that it is very different from the usual `state reduction' rule.
We also define an observable for the time-of-flight velocity   that differs, in general, from the canonical momentum observable.

\medskip

The structure of this paper in the following. In Sec. 2, we set up our notation and present the time of arrival probabilities derived by the QTP method. In Sec. 3, we study the probabilities associated to a single time-of-arrival measurement. We focus on the properties of quantum fluctuations, and we discuss possible ways to distinguish experimentally between existing proposals. In Sec. 4, we consider time-of-arrival probabilities and correlations in composite systems. In Sec. 5, we consider sequential time-of-arrival measurements. In Sec. 6, we summarize and discuss our results.

 \section{Probability assignment}
 In this section, we set-up our notation and we present the time-of-arrival probabilities derived by the QTP method. The derivation is sketched in the Appendix A. For further details, see Refs. \cite{AnSav15, AnSav12, AnSav13}.

We consider a system of non-relativistic particles described by a Hilbert space ${\cal F}$. For identical particles,  ${\cal F}$ a Fock space, either bosonic or fermionic. In the former case, ${\cal F}$ carries a representation of the canonical commutation relations
\begin{eqnarray}
[\hat{a}_{\pmb k}, \hat{a}_{\pmb k'}] = [\hat{a}_{\pmb k}^{\dagger}, \hat{a}_{\pmb k'}^{\dagger}] = 0, \hspace{1cm} [\hat{a}_{\pmb k}, \hat{a}^{\dagger}_{\pmb k'}] = \delta^3({\pmb k}-{\pmb k}'), \label{ccr}
\end{eqnarray}
expressed in terms of the bosonic annihilation and creation operators $\hat{a}_{\pmb k}$ and $\hat{a}^{\dagger}_{\pmb k}$.

 In the latter case, ${\cal F}$ carries a representation of the canonical anti-commutation relations
\begin{eqnarray}
\{\hat{c}_{\pmb k}, \hat{c}_{\pmb k'}\} = \{\hat{c}_{\pmb k}^{\dagger}, \hat{c}_{\pmb k'}^{\dagger}\} = 0, \hspace{1cm} \{\hat{c}_{\pmb k}, \hat{c}^{\dagger}_{\pmb k'}\} = \delta^3({\pmb k}-{\pmb k}'). \label{car}
\end{eqnarray}
expressed in terms of the fermionic annihilation and creation operators $\hat{c}_{\pmb k}$ and $\hat{c}^{\dagger}_{\pmb k}$.

In what follows, we will ignore spin and internal degrees of freedom of the particles, as they do not affect  time-of-arrival measurements. The quantum fields associated to the particles are
\begin{eqnarray}
 \hat{\psi}({\pmb x}) = \left\{ \begin{array}{cc} \int \frac{d^3k}{(2\pi)^{3/2}} e^{i{\pmb k} \cdot{\pmb x}} \hat{a}_{\pmb k} &
 \mbox{for bosons} \\
  \int \frac{d^3k}{(2\pi)^{3/2}} e^{i{\pmb k} \cdot{\pmb x}} \hat{c}_{\pmb k} & \mbox{for fermions}
  \end{array} \right. \label{fields}
\end{eqnarray}

For bosons, the Hamiltonian on ${\cal F} $ is $\hat{h} = \int \frac{d^3k}{(2\pi)^3} \epsilon_{\pmb k} \hat{a}^{\dagger}_{\pmb k} \hat{a}_{\pmb k}$ and for fermions $\hat{h} = \int \frac{d^3k}{(2\pi)^3} \epsilon_{\pmb k} \hat{c}^{\dagger}_{\pmb k} \hat{c}_{\pmb k}$, where $\epsilon_{\pmb k} = \frac{{\pmb k}^2}{2m}$.

\medskip

The QTP probability distribution for  a particle to be detected at time $t$  by a detector located at ${\pmb L}$ is
\begin{eqnarray}
P^{(1)}({\pmb L}, t) = C \int ds ds' \sqrt{f(t-s)f(t-s')}  \langle \Psi_0| Y^{\dagger}({\pmb L},s') \hat{Y}({\pmb L}, s) |\Psi_0\rangle, \label{plt1})
\end{eqnarray}
where $C$ is a constant, and $|\Psi_0\rangle$ the initial state of the particle system.

 The   operator $\hat{Y}({\pmb x}, s)$ is the Heisenberg-picture evolution $e^{i\hat{h}s} \hat{Y}({\pmb x}) e^{-i\hat{h}s}$ of a composite operator $\hat{Y}({\pmb x})$ that is a local functional of the quantum fields $\hat{\psi}({\pmb x})$. This operator originates from the interaction Hamiltonian between the particles and the detector. In what follows, we will consider two types of interaction,
\begin{enumerate}
\item $\hat{Y}({\pmb x}) = \hat{\psi}({\pmb x})$. This choice corresponds to a process in which the particle is absorbed during detection.
    \item $\hat{Y}({\pmb x}) = \hat{\psi}^{\dagger}({\pmb x})\hat{\psi}({\pmb x})$. This choice corresponds to a process in which the particle is scattered during detection.
\end{enumerate}

 The smearing function $f$ in Eq. (\ref{plt1}) is centered around $0$ with  a width of order $\sigma$, the temporal coarse-graining of the detector. Smearing  is essential for the definition of probabilities in the QTP method, because the time parameter $t$ is a coarse-grained quasi-classical variable that coincides with the emergence of a macroscopic record of observation in the detector.

For example, we can employ
Gaussian smearing functions
\begin{eqnarray} f(s) = \frac{1}{\sqrt{2 \pi \sigma^2}} e^{-\frac{s^2}{2\sigma^2}} \label{gauss}. \end{eqnarray}
 The Gaussians satisfy the useful identity

\begin{eqnarray}
\sqrt{f(t-s) f_{\sigma}(t-s')} = f(t - \frac{s+s'}{2}) g(s-s'), \label{eq2}
\end{eqnarray}

where

\begin{eqnarray}
g(s) = e^{-\frac{s^2}{8\sigma^2}}. \label{gsig}
 \end{eqnarray}
 The analogue of Eq. (\ref{eq2}) is also satisfied approximately for non Gaussian smearing functions. In such cases, $g$ is a positive function that    satisfies $g(0) = 1$ and $\lim_{|s|\rightarrow \infty} g(s) = 0$.

Using Eq. (\ref{eq2}), the probability distribution $P({\pmb L}, t) $ can be expressed as a convolution
\begin{eqnarray}
P^{(1)}({\pmb L}, t) = \int dt' f(t-t') P^{(1)}_{f.g.}({\pmb L}, t')
\end{eqnarray}
where
\begin{eqnarray}
P^{(1)}_{f.g.}({\pmb L}, t) = C \int d\tau  g(\tau) \langle \Psi_0| Y^{\dagger}({\pmb L},t-\frac{\tau}{2}) \hat{Y}({\pmb L}, t+\frac{\tau}{2}) |\Psi_0\rangle, \label{plt2}
\end{eqnarray}
is  a probability distribution, finer than  $P({\pmb L}, t) $, that usually takes a simple form.

The probability density associated to the measurement of one particle at time $t_1$ by a detector located at ${\pmb L}_1$ and of on particle at time $t_2$  by a detector located at ${\pmb L}_2$ is
\begin{eqnarray}
P^{(2)}({\pmb L}_1, t_1; {\pmb L}_2,t_2) = C \int ds_1ds_1'ds_2ds_2' \sqrt{f_1(t_1-s_1)f_1(t_1-s_1')f_2(t_2-s_2)f_2(t_2-s_2')}
\nonumber \\
\times \langle \Psi_0| \bar{{\cal T}}[\hat{Y}_1^{\dagger}({\pmb L}_1,s_1') \hat{Y}_2^{\dagger}({\pmb L}_2, s_2')] {\cal T}[\hat{Y}_2({\pmb L}_2, s_2) \hat{Y}_1({\pmb L}_1,s_1)]|\Psi_0\rangle, \label{pltlt}
\end{eqnarray}
where ${\cal T}$ stands for time-ordered product and $\bar{ \cal T}$ for anti-time-ordered product. Eq. (\ref{pltlt}) takes into account the possibility that each detector may be associated to a different composite operator $\hat{Y}_i({\pmb x})$.

Again $P({\pmb L}_1, t_1; {\pmb L}_2,t_2) $ can be expressed as a convolution
\begin{eqnarray}
P^{(2)}({\pmb L}_1, t_1; {\pmb L}_2,t_2) = \int dt_1' dt_2' f_1(t_1-t_1') f_2(t_2-t_2') P^{(2)}_{f.g.}({\pmb L}_1, t_1'; {\pmb L}_2,t_2'), \label{pltobs}
\end{eqnarray}
of a finer probability density
\begin{eqnarray}
P^{(2)}_{f.g.}({\pmb L}_1, t_1; {\pmb L}_2,t_2) = C \int d\tau_1d\tau_2 g(\tau_1) g(\tau_2)  \langle \Psi_0| \bar{{\cal T}}[\hat{Y}_1^{\dagger}({\pmb L}_1,t_1-\frac{\tau_1}{2}) \hat{Y}_2^{\dagger}({\pmb L}_2, t_2 - \frac{\tau_2}{2})]\nonumber \\
 \times{\cal T}[\hat{Y}_2({\pmb L}_2,  t_2 + \frac{\tau_2}{2}) \hat{Y}_1({\pmb L}_1,t_1+\frac{\tau_1}{2})]|\Psi_0\rangle. \label{pltlt2}
\end{eqnarray}
 The probability densities (\ref{plt2}) and (\ref{pltlt2}) involve averaging over the temporal and not the spatial coordinates. They have been obtained using the approximation (\ref{poscg}) in the Appendix A.2. Temporal averaging is essential for the definition of probabilities in the QTP method. This is not the case for spatial averaging. It is usually subsumed under the effects of temporal averaging and  can be omitted for simplicity.

\section{Single time-of-arrival measurement}
In this section, we revisit the time-of-arrival probability derived in Ref. \cite{AnSav12}. We examine its phase space properties, the classical limit, the associated uncertainties, and discuss how different candidates for the time-of-arrival probabilities might be distinguished experimentally.
\subsection{The ideal probability distribution}
We evaluate the fine-grained probability distribution (\ref{plt2}) for a single detection of a particle of mass $m$. We consider the simplest case of particle detection by absorption, i.e., we choose $\hat{Y}({\pmb x}) = \hat{\psi}({\pmb x})$. The resulting probability distribution is the same for fermions and bosons
\begin{eqnarray}
P^{(1)}_{f.g.} ({\pmb L}, t) = C \int \frac{d^3k d^3k'}{(2\pi)^3} \tilde{g}(\frac{\epsilon_{\pmb k}+\epsilon_{\pmb k'}}{2}) e^{i({\pmb k} - {\pmb k'})\cdot{\pmb L} - i (\epsilon_{\pmb k} - \epsilon_{\pmb k'})t} \rho_0({\pmb k},{\pmb k'}), \label{plt3}
\end{eqnarray}
where $\tilde{g}$ is the Fourier transform of the function $g$.  The function
 \begin{eqnarray}
 \rho_0^{(1)}({\pmb k},{\pmb k'})= \frac{1}{N} \langle \Psi_0|\hat{a}^{\dagger}_{\pmb k'} \hat{a}_{\pmb k}|\Psi_0\rangle
  \end{eqnarray}
  is the one-particle density matrix, where $|\Psi_0\rangle$ has been assumed a $N$-particle state.  (In Eq. (\ref{plt3}), a multiplicative factor of $N$ has been absorbed in the constant $C$.)

Let the initial state be localized  at ${\pmb x} = 0$ and the  detector  at ${\pmb x} = {\pmb L}$. We reduce the system  to one dimension along the axis that connects the particle source to the detector. The   probability density (\ref{plt3}) then becomes
\begin{eqnarray}
P^{(1)}_{f.g.} (L, t) = C \int \frac{dkdk'}{2\pi} \tilde{g}(\frac{\epsilon_{k}+\epsilon_{ k'}}{2}) e^{i( k - k')L - i (\epsilon_{k} - \epsilon_{ k'})t} \rho^{(1)}_0(k,k'). \label{plt4}
\end{eqnarray}

Eq. (\ref{plt4}) is physically meaningful only for    $t \geq 0$, but it can be mathematically extended to  $t < 0$. For initial density matrices with support only on positive momenta $k$ and  localised at $x < L$, $P_{f.g.} (L, t)$ is strongly suppressed for negative $t$. Hence, when considering the total probability of detection $\mbox{Prob}(L): = \int_0^{\infty}P_{f.g.} (L, t)  $ we can extend the range of integration to the whole real axis. Then,
 \begin{eqnarray}
 \mbox{Prob}(L)  = m C \int dk  \frac{\tilde{g}(\epsilon_k)}{|k|} \rho^{(1)}_0(k,k). \label{totdet}
 \end{eqnarray}
 The extension of integration to negative times is inadmissible for states with negative momentum, or for states with position support on both sides of the detector. Eq. (\ref{plt4}) accounts also for these cases, but Eq. (\ref{totdet}) does not apply.

 We define the absorption rate $\alpha(\epsilon)$ of the detector as the fraction of particles with incoming energy $\epsilon$ that is absorbed by the detector.
 Eq. (\ref{totdet}) implies that
 \begin{eqnarray}
 \alpha(\epsilon) \sim \tilde{g}(\epsilon)/\sqrt{2m\epsilon}.
 \end{eqnarray}
We choose the constant $C$, so that  $\mbox{Prob}(L)$ equals the fraction of detected particles. Then,
\begin{eqnarray}
P^{(1)}_{f.g.} (L, t) = \int \frac{dkdk'}{2\pi } \alpha \left(\frac{\epsilon_k + \epsilon_{k'}}{2}\right) \sqrt{\frac{\epsilon_k+\epsilon_{k'}}{m}} e^{i( k - k')L - i (\epsilon_{k} - \epsilon_{ k'})t} \rho^{(1)}_0(k,k'). \label{pltid0}
\end{eqnarray}

For a homogeneous detector of length $d << L$, the absorption rate is $\alpha(\epsilon) = \mu(\epsilon)d$, where $\mu(\epsilon)$ is the usual {\em attenuation coefficient} of the absorbing material. The attenuation coefficient can be measured directly, and it is a defining characteristic of the detector. In some cases, $\mu(\epsilon)$
can be computed from first principles as $n \sigma_{abs}$, where $n$ is the number density of the individual absorbers and $\sigma_{abs}$ is the absorption cross-section\footnote{The probability density $P^{(1)}_{f.g.} (L, t)$ of Eq. (\ref{pltid0}) is  integrated with respect to all possible loci of detection, so  it  is not a density with respect to $L$. The corresponding density can be read immediately, by substituting the absorption rate $\alpha(\epsilon)$ with the attenuation coefficient $\mu(\epsilon)$.}.

In what follows, we will consider  {\em ideal detectors}, characterized by  constant  absorption rate.  Normalizing so that $\mbox{Prob}(L)  = 1$, we obtain the {\em ideal} time-of-arrival probability distribution
\begin{eqnarray}
P^{(1)}_{id}(L,t) =  \int \frac{dkdk'}{2\pi } \sqrt{\frac{\epsilon_k+\epsilon_{k'}}{m}} e^{i( k - k')L - i (\epsilon_{k} - \epsilon_{ k'})t} \rho^{(1)}_0(k,k'). \label{pltid}
\end{eqnarray}

For initial states with  momentum spread much smaller than the mean momentum, we can approximate $\epsilon_k+\epsilon_{k'} = \frac{1}{2m}(k^2+k'^2) = \frac{1}{2m} [(k-k')^2 + 2kk'] \simeq kk'/m$. Then,  Eq. (\ref{pltid}) coincides with the time-of-arrival probability distribution of Kijowski \cite{Kij}.

The probability density (\ref{pltid}) is expressed as $Tr (\hat{\rho} \hat{\Pi}_L(t))$ where $\hat{\Pi}_L(t)$ are  positive operators with matrix elements
\begin{eqnarray}
\langle k|\hat{\Pi}_L(t)|k'\rangle = \frac{1}{2\pi } \sqrt{\frac{\epsilon_k+\epsilon_{k'}}{m}}  e^{i( k - k')L - i (\epsilon_{k} - \epsilon_{ k'})t}. \label{pilt}
\end{eqnarray}
When restricted to the subspace of states with only positive values of momentum, and when all values $t \in {\pmb R}$ are taken into account,  $\hat{\Pi}_L(t)$ defines a POVM. For general initial states, we can define a POVM by including the positive operator associated to the event of no detection
\begin{eqnarray}
\hat{\Pi}_L(\emptyset) = 1 - \int_0^{\infty} \hat{\Pi}(t) dt. \label{piempty}
\end{eqnarray}

\subsection{Time of arrival in the Wigner picture}
We bring the ideal probability distribution (\ref{pltid}) into a form that allows for a comparison with the classical time of arrival. To this end, we express the density matrix $\hat{\rho}_0^{(1)}$ in terms of its associated Wigner function
\begin{eqnarray}
W_0(X, P) = \int \frac{dy}{2\pi} \langle X-\frac{y}{2}|\hat{\rho}_0^{(1)}|X+\frac{y}{2}\rangle e^{iPy}.
\end{eqnarray}
Substituting
\begin{eqnarray}
\rho^{(1)}_0(k, k') = \frac{1}{2\pi} \int d\xi W_0(X, \frac{k+k'}{2}) e^{-iX(k-k')}.
\end{eqnarray}
into Eq. (\ref{pltid}), we obtain
\begin{eqnarray}
P^{(1)}_{id}(L,t) = \int dX dP  \frac{2P^2}{m}  u[2P(L-X - \frac{t}{m}P)] W_0(X, P), \label{pwid}
\end{eqnarray}
where
\begin{eqnarray}
u(s) := \frac{1}{2\pi} \int_{-\infty}^{\infty}  d\xi \sqrt{1+\xi^2} e^{i \xi s} = \frac{1}{\pi} \mbox{Re} \int_0^{\infty} d\xi \sqrt{1+\xi^2} e^{i\xi s} . \label{ws}
\end{eqnarray}
The integral $u(s)$, Eq. (\ref{ws}) defines a distribution function that is singular at $s = 0$. The properties of the distribution $u$ are analyzed in the Appendix B1.

  For any $a > 0$,
  \begin{eqnarray}
  u(as) = \frac{1}{2\pi a} \int_{-\infty}^{\infty}  dy \sqrt{1+ (y/a)^2} e^{i y s}.
  \end{eqnarray}

   For  large values of $a$, we can expand  the square root, to obtain a formal series
\begin{eqnarray}
u(as) = \frac{1}{a} \delta (s) - \frac{1}{2a^3} \delta''(s) - \frac{1}{8a^5} \delta''''(s) + \ldots  \label{uexpand}
\end{eqnarray}

We use Eq. (\ref{uexpand}) in order to express the probability density Eq. (\ref{pwid}) as a series
\begin{eqnarray}
P^{(1)}_{id}(L,t)  = P^{(1)}_{cl}(L,t) + P^{(1)}_1(L, t) + P^{(2)}_2(L, t) + \ldots
\end{eqnarray}
The first term in the series
\begin{eqnarray}
P^{(1)}_{cl}(L, t) &=& \int dXdP \frac{|P|}{m} \delta(L-X - \frac{P}{m}t) W_0(X, P)
\nonumber \\
 &=&   \int dXdP \delta (t - m\frac{L-X}{P}) W_0(X, P) \label{pcl}
\end{eqnarray}
coincides with   the probability distribution associated to the {\em classical} time-of-arrival observable \cite{Werner}
\begin{eqnarray}
T_{c}(X, P) = \frac{m(L-X)}{P}.
\end{eqnarray}

The associated operator
\begin{eqnarray}
\hat{T}_{c} = \frac{1}{2} [(L-\hat{x})\hat{p}^{-1}+\hat{p}^{-1}(L-\hat{x})]
 \end{eqnarray}
 was first studied by Aharonov and Bohm \cite{AhBo61}. $\hat{T}_c$ is Hermitian  but not self-adjoint. However, when restricted to states with support on positive momentum and well localised at positions $x < L$, $\hat{T}_c$ is indistinguishable from its self-adjoint variants \cite{DeMu97, MaIs}.

The action of $\hat{T}_c$ on  states $|\psi\rangle$ with support on strictly positive momenta is well defined. For such  states, $\hat{T}_{c}$ and the Hamiltonian $\hat{H}$ satisfy a canonical commutation relation.
\begin{eqnarray}
[\hat{T}_{c} , \hat{H}] |\psi\rangle = - i |\psi\rangle. \label{cctime}
\end{eqnarray}

The first correction to the classical distribution is
\begin{eqnarray}
P_1^{(1)}(L, t) = - \frac{1}{8m}\int  dP \frac{\partial_X^2W_0(L- \frac{P}{m}t, P)}{|P|}.
\end{eqnarray}
This term diverges, unless the Wigner function   vanishes at $P = 0$.

%The Wigner function $W_0(X, P)$ may take negative values, so  it does not correspond to a classical probability distribution. There is no guarantee that $P^{(1)}_{cl}(t)$ is positive definite, and hence, a genuine probability distribution. The function $P_{cl}(t)$ is obtained as an approximation to the genuine probability distribution Eq. (\ref{pwid}), and it is expected to be well-defined only in the regime where the approximation holds.

\subsection{Non-classical effects}
The moment-generating function of the probability distribution (\ref{pwid}) is
\begin{eqnarray}
Z^{(1)}[\mu] := \int_{-\infty}^{\infty} dt P^{(1)}_{id}(t) e^{-i \mu t} = \langle e^{-i \mu T_c} \sqrt{1 + \frac{\mu^2}{16H^2}} \rangle \label{zmu}
\end{eqnarray}
where $H(X, P) = \frac{P^2}{2m}$, and  we wrote
\begin{eqnarray}
\langle F \rangle = \int dXdP F(X, P) W_0(X, P)
\end{eqnarray}
in order to denote averaging with respect to the Wigner function $W_0$.

For $\langle\mu^2/H^2\rangle << 1$,  $Z^{(1)}[\mu]$ is well approximated by   the generating function of the classical observable $T_c$. Thus,   $P^{(1)}_{id}(L,t)  \simeq  P^{(1)}_{cl}(L,t)$, except for the regime of very low kinetic energies or  very early times (large $\mu$).

The expectation  value  $\bar{t}$  and the  mean deviation $\Delta t$ of the time of arrival are
\begin{eqnarray}
\bar{t} &=& \langle T_c\rangle \label{ehren}\\
(\Delta t)^2 &=& (\Delta T_c)^2 - \frac{1}{16} \langle H^{-2}\rangle. \label{devt}
\end{eqnarray}

For states with support on strictly positive momenta, Eq. (\ref{cctime}) implies the uncertainty relation  $\Delta T_c \Delta H \geq \frac{1}{4}$. Then, Eq. (\ref{devt}) becomes
\begin{eqnarray}
(\Delta t)^2 \geq \frac{1}{4(\Delta H)^2} - \frac{1}{16} \langle H^{-2}\rangle. \label{uncrel}
\end{eqnarray}
The lower bound to $\Delta t$  is smaller by what one would  surmise from a naive application the Kennard-Robertson inequality to the time-of-arrival operator

The analogue of Eq. (\ref{uncrel}) for Kijowski's POVM  has a plus sign in front of the $\langle H^{-2}\rangle$ term, and thus, implies that  $(\Delta t) \geq \frac{1}{4} \sqrt{\langle H^{-2}\rangle}$. By Jensen's inequality,  $\langle H^{-2}\rangle \geq \langle H\rangle^{-2}$, and    an uncertainty relation    $\langle H\rangle \Delta t \geq \frac{1}{4}$ follows. This is of the same form (modulo a constant of order unity) with the inequality derived in Ref. \cite{BSPME}. However, {\em no such  uncertainty relation  exists for the POVM (\ref{pltid})}.

 %However, we have to keep in mind that the time-of-arrival observable, as defined by the QTP method, is coarse-grained at a scale $\sigma$. If $\Delta t$ is significantly smaller than $\sigma$, quantum fluctuations are unobservable.

Two types of non-classical effects are manifested in the probability distribution  Eq. (\ref{pwid}). First, the classical time-of-arrival observable may exhibit quantum interference, as a consequence of the non-classical character of the initial state.
In this case, the time of arrival behaves like any other phase space variable. An oscillating behavior of the Wigner function $W_0$ in some region of the phase space leads to interference terms in the probability distribution. For example, we consider a superposition state $c_1 |\phi_1\rangle + c_2 |\phi_2\rangle$, where $|\phi_1\rangle$ corresponds to a Wigner function localized at $(X_1, P_1)$ and
$|\phi_2\rangle$ corresponds to a Wigner function localized at $(X_2, P_2)$. Then, the probability distribution for the time of arrival exhibits two peaks at $t_1 = T_{cl}(X_1, P_1)$ and $t_2 =  T_{cl}(X_2, P_2)$ and   by oscillatory terms in the intermediate values of $t$.

 The other  non-classical effect is that the time-of-arrival probability $P^{(1)}_{id}$ may differ significantly  from the probability $P^{(1)}_{cl}$ that is defined in terms of the classical observable $T_c$.   The difference between the two distributions is significant if $\Delta t \sqrt{\langle H^{-2}\rangle}$ is of order unity or smaller, and it is negligible if  $\Delta t /\sqrt{\langle H^{-2}\rangle} >> 1$. The latter condition is satisfied if
 \begin{eqnarray}
 \Delta t \langle H\rangle >>1. \label{valid1}
 \end{eqnarray}
 Eq. (\ref{valid1}) is a classicality condition for the time of arrival \footnote{
 Conditions  similar to Eq. (\ref{valid1}) have appeared in several approaches to the time of arrival based upon measurement models and/or complex potentials \cite{AOPRU, ECM08, HaYe09, YDHH11}. In those models, $\Delta t$ is not the mean deviation of the probability distribution, but the accuracy in the determination of the time of arrival. This quantity coincides with the temporal coarse-graining parameter $\sigma$ that is introduced in the QTP method---see, Sec. 2. As a matter of fact, the first derivation of the probability density (\ref{pltid}) required the hypothesis that $\sigma \langle H\rangle >>1$ \cite{AnSav06}. This condition was not necessary in later derivations that described the interaction between quantum particle and measurement apparatus   in terms of local quantum fields \cite{AnSav12}.

The coarse-graining scale $\sigma$ appears as a parameter in the absorption rate $\alpha(\epsilon)$ of the detector, and, as such, enters the probability distribution (\ref{pltid0}). The deviation from the ideal distribution (\ref{pltid}) is $\sigma$-dependent; however, there is no {\em a priori} reason, why this dependence is stronger at the low momentum limit.
In our opinion, the requirement that $\sigma \langle H\rangle >> 1$  is not a fundamental condition upon the measurability of the time of arrival.}.

\subsection{Distinguishing between different time-of-arrival proposals}
While all physically reasonable proposals for the time-of-arrival probability have to coincide at the classical limit, they are expected to differ in their description of non-classical effects.

In order to avoid complications inessential to the main argument, we restrict to states with support on positive values of momentum, and localized at $x < L$, so that  the probability of no detection is negligible. Hence, the probability density for the time of arrival is normalized to unity.

We will consider  ideal probability distributions  $P^{(1)}(L, t)$ that do  not depend on any parameters that characterize the measuring apparatus.  We  assume that for large momenta,  $P^{(1)}(L, t)$  the classical time-of-arrival variable $T_c$. These conditions imply that the moment-generating function  $Z^{(1)}[\mu] := \int_{-\infty}^{\infty} dt P^{(1)}(L, t) e^{-i \mu t}$ has the  form
\begin{eqnarray}
Z^{(1)}[\mu] = \langle e^{-i\mu T_c} \eta(\mu/H)\rangle \label{zmu2}
\end{eqnarray}
for some positive function $\eta(x)$ of the dimensionless quantity $x = \mu/H$. The function $\eta$ depends only on the ratio $\mu/H$, because $\mu$ has the dimensions of energy, and in the absence of other parameters with dimension of energy (characterizing the apparatus), $\mu/H$ is the only possible combination.

We require that $\eta(x)$ satisfies the following properties.

\begin{enumerate}[(i)]
\item $\eta(0) = 1$, since $P^{(1)}(L, t)$ is normalized to unity.
\item $\eta'(0) = 0$, so that Eq. (\ref{ehren}) holds. This condition guarantees that the expectation value always coincide with that of the classical observable $T_c$. Eq. (\ref{ehren}) is the analogue of Ehrenfest's theorem for the time of arrival.
 \item $\eta(-x) = \eta(x)$, so that the time-of-arrival probabilities are time-reversal covariant.
\end{enumerate}

%This difference is reflected in all moments of the distribution. For example, the mean deviation is
%\begin{eqnarray}
%(\Delta t)^2 = (\Delta T_c)^2 - f''(0) \langle H^{-2}\rangle. \label{devt2}
%\end{eqnarray}

From Eq. (\ref{zmu2}), we express the probability density $P^{(1)}(L, t)$ in terms of the density matrix $\hat{\rho}^{(1)}$,
\begin{eqnarray}
P^{(1)}(L, t) = \int \frac{dkdk'}{2\pi } \frac{k+k'}{2m}\eta\left(\frac{4(k'-k)}{k+k'}\right) e^{i( k - k')L - i (\epsilon_{k} - \epsilon_{ k'})t} \rho^{(1)}_0(k,k'). \label{pltidd}
\end{eqnarray}

The probability densities of the form  (\ref{pltidd}) transform covariantly under time translations. They are special cases of Werner's   time-of-arrival probability distribution  \cite{Werner}.

Different choices of the function $\eta$ correspond to different proposals for an ideal time-of-arrival probability distribution.

\begin{enumerate}[(i)]
\item The probability density (\ref{pltid}) defined through the QTP method corresponds to $\eta(x) = \sqrt{1+\frac{x^2}{16}}$.
\item Kijowski's probability distribution \cite{Kij} corresponds to $\eta(x) = \sqrt{1-\frac{x^2}{16}}$.
\item Several different proposals  correspond to $\eta(x) = 1$. Proposals based on defining self-adjoint variants of the operator $\hat{T}_c$ \cite{DeMu97, MaIs}, lead to the probability density $P^{(1)}_{cl}(L, t)$ when restricted to states with positive momentum and $x < L$. This is also the case for the time-of-arrival probability defined by the probability current \cite{ML} and for some measurement models \cite{BEMS,  DEHM}.
\end{enumerate}

 All probability densities (\ref{pltidd}) have the same behavior at high momentum. This is also true for some proposed time-of-arrival probabilities that are non-covariant with respect to time translations \cite{HELM}. Thus, the different proposals can only be distinguished by their predictions in the low momentum regime.   However, moments such as $ (\Delta T_c)^2$ and $\langle H^{-2}\rangle$  cannot be used for this purpose because they diverge.

We quantify the low momentum behavior of the time-of-arrival probabilities  in terms of a temperature variable. We consider  an initial state with a thermal distribution  of positive momenta, at temperature $\beta^{-1}$
\begin{eqnarray}
W_0(X,  P) = n_0(X) \sqrt{\frac{2\beta}{\pi m}} e^{-\frac{1}{2}\beta P^2/m} \theta(P);
\end{eqnarray}
$n_0$ is a probability distribution for position  with zero mean and mean deviation $\sigma_X$. For $\sigma_X << L$,  we can  substitute $n_0$ with a delta function. The   probability density $P^{(1)}(L, t)$ turns out to be of the form $\int_0^{\infty} d x \eta(x) G(x)$, for some function $G(x)$. The exact form of $G(x)$ is not relevant, only the fact that $G(x)$  decays as $e^{-2mL^2x^2/\beta}$ for large $x$. If $mL^2/\beta$ is significantly larger than unity, only values of $x$ very close to zero contribute to the integral, and different functions $\eta(x)$ lead to the same probability density. Hence, a distinction between different candidate  time-of-arrival probabilities is possible only if
the quantity
\begin{eqnarray}
\nu = \frac{\beta}{mL^2}
\end{eqnarray}
is at least of   order one. This means that the {\em distance $L$ between source and detector must be of the   order of the thermal de Broglie wave-length of the particles}.

 We also have to take into account that the temporal scale $\tau =  L\sqrt{m \beta}$ for the measured time of arrival  must be significantly larger than the time resolution of the apparatus. Lowering the  temperature increases both $\nu$ and $\tau$. For $\beta^{-1}$ near a tenth of a milli-Kelvin, $\nu $ becomes of order unity, while $\tau \sim 1 \mu s$ is significantly larger than typical resolution  of solid state detectors. Such a measurement requires $L \sim 30 \mu m$ for electrons and $L\sim 1\mu m$ for neutrons.

 These estimates suggest that the quantum regime above is not beyond present capabilities.   Thus, an experimental distinction between different proposals for the time-of-arrival is  possible in principle, even if the realization of such an experiment may very difficult.   Perhaps, the distinction can be made easier in a different set-up,   namely, time-of-arrival measurements in multi-partite systems that are discussed in the following section.

\section{Time of arrival in composite systems}
The QTP method also applies to set-ups that involve  more than one time-of-arrival measurement.  In this section, we study time of arrival measurements in a bipartite system.
\subsection{Probability assignment}
We  evaluate the fine grained probability distribution (\ref{pltlt2}) for two detection events. We assume that  both particles are absorbed at detection, i.e., $\hat{Y}_1({\pmb x}) = \hat{Y}_2({\pmb x})= \hat{\psi}({\pmb x})$. Then,
\begin{eqnarray}
P^{(2)}_{f.g.}({\pmb L}_1, t_1; {\pmb L}_2,t_2) = C \int d^3k_1 d^3k_2 d^3k_1' d^3k_2'  \tilde{g}(\frac{\epsilon_{{\pmb k}_1}+\epsilon_{{\pmb k'}_1}}{2})\tilde{g}(\frac{\epsilon_{{\pmb k}_2}+\epsilon_{{\pmb k'}_2}}{2})
\nonumber \\
\times e^{i({\pmb k}_1 - {\pmb k'}_1) \cdot {\pmb L}_1 - i (\epsilon_{{\pmb k}_1} - \epsilon_{{\pmb k'}_1})t_1}   e^{i({\pmb k}_2 - {\pmb k'}_2) \cdot {\pmb L}_2 - i (\epsilon_{{\pmb k}_2} - \epsilon_{{\pmb k'}_2})t_2} \rho_0^{(2)}({\pmb k}_1, {\pmb k}_2|{\pmb k}_1', {\pmb k}_2'),
\label{pltlt2b}
\end{eqnarray}
where
\begin{eqnarray}
\rho_0^{(2)}({\pmb k}_1, {\pmb k}_2|{\pmb k}_1', {\pmb k}_2') = \frac{1}{N^2-N}\langle \Psi_0|\hat{c}^{\dagger}_{{\pmb k'}_1} \hat{c}^{\dagger}_{{\pmb k'}_2} \hat{c}_{{\pmb k}_2} \hat{c}_{{\pmb k}_1} |\Psi_0\rangle,
\end{eqnarray}
 and $|\Psi_0\rangle$ was assumed to be a $N$-particle state. Eq. (\ref{pltlt2b}) applies for both bosons and fermions, the only difference being the symmetrization properties of the density matrix $\hat{\rho}^{(2)}_0$.

We consider a particle source  localized at ${\pmb x} = 0$ and    two detectors   localized at ${\pmb x} = {\pmb L}_1$ and ${\pmb x}_2 = {\pmb L}_2$, respectively. For  distances $L_i = |{\pmb L}_i|$  much larger than the dimensions of the source, we can treat  each particle as one dimensional, moving along the axis connecting the source to the detector. Following the procedure of Sec. 3.1, we obtain an ideal probability distribution that generalizes Eq. (\ref{pltid}),
\begin{eqnarray}
P^{(2)}_{id}(L_1, t_1; L_2, t_2) =   \int \frac{dk_1dk_1'dk_2 dk_2'}{4\pi^2 m^2} \sqrt{\frac{\epsilon_{k_1}+\epsilon_{k_1'}}{m}} \sqrt{\frac{\epsilon_{k_2}+\epsilon_{k_2'}}{m}} \nonumber \\
\times e^{i( k_1 - k_1')L_1 - i (\epsilon_{k_1} - \epsilon_{ k_1'})t_1} e^{i( k_2 - k_2')L_2 - i (\epsilon_{k_2} - \epsilon_{ k_2'})t_2} \rho^{(2)}_0(k_1,k_2;,k_1',k_2') \label{pltltid}
\end{eqnarray}

Eq. (\ref{pltltid}) can be written as
\begin{eqnarray}
P^{(2)}_{id}(L_1, t_1; L_2, t_2)  = Tr \left[ \hat{\rho}^{(2)}_0 \hat{\Pi}_{L_1}(t_1) \otimes \hat{\Pi}_{L_2}(t_2)\right], \label{prb2}
\end{eqnarray}
where the positive operators $\hat{\Pi}_L(t)$ are given by Eq. (\ref{pilt}).

It is straightforward to express  the probability density (\ref{prb2}) in terms of a Wigner function $W_0(X_1, X_2, P_1, P_2)$ for a pair of particles
\begin{eqnarray}
 P^{(2)}_{id}(L_1, t_1; L_2, t_2)  = \int d^2X d^2 P \frac{4P_1^2P_2^2}{m^2}  u[2P_1(L_1-X_1 - \frac{t_1}{m}P_1)]
  \nonumber \\
\times
 u[2P_2(L_2-X_2 - \frac{t_2}{m}P_2)]
 W_0(X_1, X_2, P_1, P_2).  \label{prb2b}
\end{eqnarray}
The  corresponding moment-generating function is
\begin{eqnarray}
Z^{(2)}[\mu_1, \mu_2]:=  \int dt_1 dt_2 P^{(2)}_{id}(L_1, t_1; L_2, t_2) e^{-i \mu_1 t_1 - i \mu_2 t_2} \nonumber \\
= \langle e^{-i \mu_1 T_{c1}- i \mu_2 T_{c2}}\sqrt{1+ \frac{\mu_1^2}{16H_1^2}}\sqrt{1+ \frac{\mu_2^2}{16H_2^2}}\rangle. \label{z2}
\end{eqnarray}

The probability density (\ref{prb2}) has to be supplemented with probabilities for  the events of no detection in either detector, namely,
\begin{eqnarray}
P^{(2)}_{id}(L_1, \emptyset; L_2, t_2) = Tr \left[ \hat{\rho}_0 \hat{\Pi}_{L_1}(\emptyset) \otimes \hat{\Pi}_{L_2}(t_2)\right] \label{p2emptya}\\
P^{(2)}_{id}(L_1, t_1; L_2, \emptyset)  = Tr \left[ \hat{\rho}_0 \hat{\Pi}_{L_1}(t_1) \otimes \hat{\Pi}_{L_2}(\emptyset)\right], \\
\mbox{Prob}^{(2)}_{id}(L_1, \emptyset; L_2, \emptyset)  = Tr \left[ \hat{\rho}_0 \hat{\Pi}_{L_1}(\emptyset) \otimes \hat{\Pi}_{L_2}(\emptyset)\right]. \label{p2emptyc}
\end{eqnarray}

Using Eq. (\ref{piempty}) we obtain a relation between $P^{(2)}$ and $P^{(1)}$
\begin{eqnarray}
\int_0^{\infty} dt_2 P^{(2)}_{id}(L_1, t_1; L_2, t_2)  + P^{(2)}_{id}(L_1, t_1; L_2, \emptyset) = P^{(1)}(L_1, t_1).
\end{eqnarray}

As in Sec. 3.1, the probabilities (\ref{p2emptya}---\ref{p2emptyc}) vanish for initial states with support only on positive values of momenta. In this case,
\begin{eqnarray}
\int_0^{\infty} dt_2 P^{(2)}_{id}(L_1, t_1; L_2, t_2)   = P^{(1)}(L_1, t_1). \label{redux1}
\end{eqnarray}

 Eq. (\ref{prb2}) also applies  for pairs of distinguishable particles. The derivation requires the  use of  a different field for each type of particle, otherwise, it proceeds in exactly the same way. The  differences are that (i)
 the initial density matrix $\hat{\rho}_0^{(2)}$ is not restricted to the antisymmetric or symmetric subspace of the particle's Hilbert space and (ii) that the two particles may have different masses.

\subsection{Incompatibility with probability currents}

The use of probability currents is the oldest, and arguably the simplest, approach to the time-of-arrival problem.
The time-of-arrival probability density for a single particle is the expectation value of a  current  operator $\hat{J}(L,t)$ as
    $ \langle \psi_0|\hat{J}( L, t)|\psi_0\rangle$ on an initial state
  $|\psi_0\rangle$.
The usual probability current of Schr\"odinger's equation corresponds to
\begin{eqnarray}
\hat{J}(L,t) = e^{i \hat{H}t}[ \hat{p}\delta(\hat{x}-L) + \delta(\hat{x}-L)\hat{p}] e^{-i \hat{H}t}. \label{cop0}
 \end{eqnarray}
 This  choice for $\hat{J}(L,t)$ is not satisfactory, because it does not lead to positive definite probabilities \cite{Allcock, BrMe}. However, the probability densities obtained from POVMs can be expressed as operator-ordered variations of the Schr\"odinger current. For example, Kijowski's probabilities correspond to a current operator
 \begin{eqnarray}
 \hat{J}(L,t) = e^{i \hat{H}t}|\hat{p}|^{1/2}\delta(\hat{x}-L)  |\hat{p}|^{1/2}e^{-i \hat{H}t}.
\end{eqnarray}
Thus, for a single particle, approaches based on the notion of a probability current do not lead to significantly different predictions from the time-of-arrival POVMs. This equivalence fails   in multi-partite systems. Time-of-arrival probabilities defined in terms of probability currents should be of the form
\begin{eqnarray}
P^{(1)}( L, t)  &=&   \langle \Psi_0|\hat{J}( L, t)|\Psi_0\rangle\label{p1cur} \\
 P^{(2)}(L_1, t_1; L_2, t_2) &=& \langle \Psi_0| \hat{J}(L_1, t_1)\hat{J}(L_2, t_2)|\Psi_0\rangle, \label{p2cur}
\end{eqnarray}
for $|\Psi_0\rangle \in {\cal F}$ some current operator $\hat{J}(L, t)$ on ${\cal F}$.  The current operator  should satisfy $[\hat{J}( L_1, t_1), \hat{J}(L_2, t_2)] = 0$, so that the probability (\ref{p2cur}) is real-valued. By "current operator" we mean any operator on ${\cal F}$ that depends on $L$ and $t$ and that defines positive definite probabilities. When restricted to the one-particle subspace, it should correspond to the standard probability current, Eq. (\ref{cop0}), modulo operator-ordering.

For systems of identical particles,  Eqs. (\ref{p1cur}---\ref{p2cur}) are not compatible with Eqs. (\ref{pltid}) and (\ref{p2emptyc}). To see this, consider a two-particle state $|\Psi_0 \rangle$.  Eq. (\ref{p1cur}) reproduces the single particle distribution corresponding to a POVM $\hat{\Pi}_{ L}(t)$, if it is of the form
\begin{eqnarray}
\hat{J}(L,t) = \frac{1}{2} \left(\hat{\Pi}_{ L}(t) \otimes \hat{1} + \hat{1} \otimes \hat{\Pi}_{ L}(t)\right), \label{cur1}
\end{eqnarray}
where we
  have taken into account that any physical operator must be invariant under exchange of the two identical particles.

Comparing Eq. (\ref{p2cur}) and Eq. (\ref{prb2}), we obtain
\begin{eqnarray}
\hat{J}(L_1, t_1)\hat{J}({ L}_2, t_2) =  \hat{\Pi}_{L_1}(t_1) \otimes \hat{\Pi}_{L_2}(t_2) \label{cur2}
\end{eqnarray}
Eqs. (\ref{cur1}) and (\ref{cur2}) are clearly incompatible.  The time-of-arrival probabilities obtained by the QTP method cannot be expressed in terms of a probability current. In fact, any approach to the time-of-arrival in terms of POVMs would lead to an equation similar to (\ref{prb2}) when applied to composite systems. We conclude that {\em current-based approaches strongly disagree with POVM-based approaches} in multi-partite systems.

\medskip

Probabilities defined in terms of a current operator are subject to constraints that do not apply to probabilities defined through POVMs. To prove this, we  first define
  the two-time {\em coherence function}
\begin{eqnarray}
C^{(2)}(L_1, t_1; L_2, t_2) = \frac{P^{(2)}(L_2, t_2; L_1, t_1)}{ P^{(1)}(L_1, t_1) P^{(1)}(L_2, t_2)},  \label{coha}
\end{eqnarray}
where the probability distributions $P^{(2)}(L_2, t_2; L_1, t_1)$ and $P^{(1)}(L, t) $ satisfy Eq. (\ref{redux1}).

%First, by applying the Cauchy-Schwartz inequality on Eq. (\ref{p2cur}), we obtain $P^{(2)}(L_2, t_2; L_1, t_1) \leq \sqrt{P^{(2)}(L_2, t_2; L_2, t_2) P^{(2)}(L_1, t_1; L_1, t_1)  }$. By Eq. (\ref{coha}), we derive the inequality
%\begin{eqnarray}
%C^{(2)}(L_1, t_1; L_2, t_2) \leq \sqrt{C^{(2)}(L_1, t_1; L_1, t_1)C^{(2)}(L_2, t_2; L_2, t_2)}. \label{bunching}
%\end{eqnarray}

The diagonal elements of $C^{(2)}$ define the {\em coincidence function},
\begin{eqnarray}
c^{(2)}(L,t):= C^{(2)}(L, t; L, t)  \label{superp0}
\end{eqnarray}
  For $c^{(2)}(L,t) > 1$ simultaneous detection is more probable than what would be predicted if the events were statistical independent, while for $c^{(2)}(L,t) < 1$ simultaneous detection is less probable.

  %Thus, for time-of-arrival measurements, the condition $c^{(2)}(L,t) > 1$ describes particle bunching, and the condition $c^{(2)}(L,t) < 1$ particle anti-bunching\footnote{We use a different terminology from quantum optics, where the  condition  $c^{(2)}(L,t) > 1$ defines "super-Poissonian photon statistics", and the word "bunching" refers to a different condition \cite{WM}. We have chosen different terms, because the  differences between photons and non-relativistic particles  make the  direct translation of photon concepts non-intuitive. Unlike photons, massive particles do not exhibit superpositions of different particle number.  Hence, notions related to the distribution of particle numbers do not make sense.}

For the probability densities (\ref{p1cur}---\ref{p2cur}), the coincidence function satisfies
\begin{eqnarray}
c^{(2)}(L,t) = \frac{ \langle   \hat{J}(L, t)^2\rangle}{\langle \hat{J}(L, t)\rangle^2} \geq 1 + \left(\frac{\Delta J(L,t)}{\langle \hat{J}(L,t)\rangle}\right)^2 \geq 1, \label{superp}
\end{eqnarray}
where $\Delta J(L, t)$ stands for the standard deviation of    $\hat{J}(L,t)$.  Thus, any measurement of the coincidence function  that violates Eq. (\ref{superp})  for some value of $t$ disproves the definition of time-of-arrival probabilities in terms of probability currents.

 In Sec. 4.4, we present  explicit examples of coincidence functions  that  violate  Eq. (\ref{superp}) very strongly. For example, for some fermionic quantum states, $c^{(2)}(L,t)$ may vanish. These examples strongly  suggest that the violation of (\ref{superp}) is measurable, and so is the distinction between POVM-based and current-based approaches to the time-of-arrival problem.

 We note that the QTP method leads not only to a probability density of the form (\ref{prb2}), but also to the specific expression (\ref{pilt}) for the POVM (\ref{pilt}).  As shown above, the predictions of the QTP method can sharply be distinguished  from those of current-based theories. The distinction from other POVM-based approaches, for example,  POVMs $\hat{\Pi}_L(t)$ other than (\ref{pilt}), is rather more difficult. It would require exploring the regime of low momenta, as discussed in Sec. 3.4.

%The analogue of Eq. (\ref{bunching}) in quantum optics is the statement that photons are detected in bunches ({\em photon bunching}), while the analogue of Eq. (\ref{superp}) is that photons satisfy super-Poissonian statistics. Both inequalities may be violated for photons.
%The interpretation of the quantities defined above is rather different In the time-of-arrival context. The reasons lie in the fundamental differences of massive particles from photons. First, massive particles may move at different speeds, while photons move at the speed of light. Second, there exist superposition states of different photon number, while this is usually not the case for massive particles.

%We conclude that any theory  that views probability currents as essential for the definition of time-of-arrival probabilities leads to very different predictions for measurements in multi-partite systems, when compared to theories based on POVMs.

\subsection{Non-classical correlations}
Consider  a bipartite particle system described   by a {\em classical} probability density  $\rho_0(\xi)$ on a state space $\Gamma$.   The joint probability for two time-of-arrival measurements is of the form
\begin{eqnarray}
P^{(2)}(L_1, t_1; L_2, t_2) = \int d \xi \rho_0(\xi) F_{L_1,t_1}(\xi) F_{L_2, t_2}(\xi), \label{cl2lt}
\end{eqnarray}
where $F_{L, t}$ are positive valued functions on $\Gamma$.   The classical probability density
 (\ref{cl2lt}) is subject to constraints that do not limit the quantum probability density (\ref{prb2b}). We shall express some of these constraints in terms of two   inequalities for the coherence function (\ref{coha}).

 The first constraint follows from the Cauchy-Schwartz inequality for Eq. (\ref{p2cur}),
 \begin{eqnarray}
 P^{(2)}(L_2, t_2; L_1, t_1) \leq \sqrt{P^{(2)}(L_2, t_2; L_2, t_2) P^{(2)}(L_1, t_1; L_1, t_1)  },
 \end{eqnarray}
 or equivalently,
\begin{eqnarray}
C^{(2)}(L_1, t_1; L_2, t_2) \leq \sqrt{c^{(2)}(L_1,t_1)c^{(2)}(L_2,t_2)}. \label{bunching}
\end{eqnarray}

We shall refer to Eq. (\ref{bunching}) as the {\em C-S constraint}.
%For $L_1 = L_2 = L$, Eq. (\ref{bunching}) implies that
%\begin{eqnarray}
%h(t_1, t_2):= \frac{C^{(2)}(L, t_1; L, t_2) }{\sqrt{c^{(2)}(L,t_1)c^{(2)}(L,t_2) }} \leq 1. \label{hratio}
%\end{eqnarray}

The second constraint is formally identical with Eq. (\ref{superp}),
\begin{eqnarray}
c^{(2)}(L,t) = \frac{ \langle   F_{L, t}^2\rangle}{\langle F_{L, t}\rangle^2}   \geq 1, \label{superp2}
\end{eqnarray}
where here $\langle\ldots\rangle$ stands for averaging with respect to $\rho_0$.

We shall refer  to Eq. (\ref{superp}) as the {\em bunching} condition, meaning that the simultaneous detection of particles is enhanced. The condition $c^{(2)}(L,t)  < 1$ will be referred to as {\em antibunching}\footnote{We decided to use a different terminology from quantum optics, where bunching refers to the analogue of Eq. (\ref{bunching}) (for stationary or almost stationary states) and condition (\ref{superp}) is referred to as "super-Poissonian photon statistics" \cite{WM}.  Unlike photons,  massive particles do not exhibit superpositions of different particle number.  Hence, notions related to the distribution of particle numbers are not useful.}. Both constraints can be violated by systems prepared in an entangled state---some examples are provided in Sec. 4.4.  Therefore,  {\em time-of-arrival measurements define entanglement witnesses}.  The relation between time of arrival, entanglement and other quantum resources will be elaborated in other publications.

Here we restrict to a simple example that demonstrates  how entanglement information is encoded into   time-of-arrival correlations. For simplicity, we will consider a system of distinguishable particles, in order to avoid  subtleties in the definition of separability in systems of identical particles. We characterize separability using the Peres-Horodecki criterion \cite{PPT} for continuous variables \cite{Simon}.

We restrict to states $\hat{\rho}_0^{(2)}$ with strictly positive momentum content for both particles, so that the action of the operators $\hat{T}_{ci}$ is well defined. Then we define the operators
 $\hat{T}_\pm = \hat{T}_{c1}\pm \hat{T}_{c2}$ and $\hat{H}_{\pm} = \hat{H}_1 \pm \hat{H}_2$. For any separable state, the following inequalities hold
\begin{eqnarray}
\Delta H_+ \Delta T_- \geq 1 \hspace{0.8cm} \Delta H_- \Delta T_+ \geq 1.
\end{eqnarray}

We evaluate the mean deviation of the variables, $t_{\pm} = t_1\pm t_2$ with respect to the probability distribution (\ref{prb2b})
\begin{eqnarray}
(\Delta t_{\pm})^2 = (\Delta T_{\pm})^2 - \frac{1}{16} \langle H_1^{-2}+H_2^{-2}\rangle.
\end{eqnarray}
 Thus, for a separable initial state, the following inequalities are satisfied
 \begin{eqnarray}
 (\Delta t_\pm)^2 \geq \frac{1}{(\Delta H\mp)^2} - \frac{1}{16} \langle H_1^{-2}+H_2^{-2}\rangle. \label{ineq+-}
 \end{eqnarray}
 Conversely, if Eq. (\ref{ineq+-})  is violated, the state of the system is entangled.

 We specialize to the case of an initial density matrix that is symmetric under particle exchange. Then, $(\Delta t_1)^2 = (\Delta t_2)^2 = (\Delta t)^2$, $(\Delta H_1)^2 = (\Delta H_2)^2 = (\Delta H)^2$, and $\langle H_1^2\rangle = \langle H_2^2\rangle = \langle H^2\rangle$. For this state, $(\Delta t_{\pm})^2 = 2 (\Delta t)^2 \pm 2 C_{t_1t_2}$ and  $(\Delta H_{\pm})^2 = 2 (\Delta H)^2 \pm 2 C_{H_1H_2}$, where $C_{AB} = \langle AB\rangle - \langle A\rangle \langle B\rangle$ is the correlation function of the observables $A$ and $B$. Then, the separability conditions (\ref{ineq+-}) become
\begin{eqnarray}
(\Delta t)^2 \pm C_{t_1t_2} \geq \frac{1}{4((\Delta H)^2 \mp C_{H_1H_2})} - \frac{1}{16} \langle H^{-2}\rangle.
\end{eqnarray}

\subsection{An example}
We  evaluate the probability density (\ref{prb2}) for  states of the form
\begin{eqnarray}
|\Psi_0\rangle = \frac{1}{\sqrt{2}}\left(|\psi_1\rangle \otimes |\psi_2\rangle \pm |\psi_2\rangle \otimes |\psi_1\rangle \right), \label{psi02}
\end{eqnarray}
for two orthogonal single-particle states $|\psi_1\rangle$ and $|\psi_2\rangle$. The plus sign in Eq. (\ref{psi02}) corresponds to bosonic and the minus sign to fermionic particles.

We assume that  $|\psi_1\rangle$ has support   at  large  momenta, so that the corresponding
 single-time  probability densities   (\ref{pltidd}) are indistinguishable; we assume the same for $|\psi_2 \rangle$. Then we can approximate
   $\hat{\Pi}_L(t)$ with   Kijowski's POVM, so that $\langle \psi_j|\hat{\Pi}_L(t)|\psi_i\rangle = a_i(L,t) a_j^*(L,t)$,
   where
 \begin{eqnarray}
 a_i(L, t) = \int \frac{dk}{\sqrt{2 \pi m}} \sqrt{|k|} e^{ikL - i \epsilon_kt} \tilde{\psi}_i(k).
 \end{eqnarray}

Then, we obtain
 \begin{eqnarray}
 P^{(1)}_{id}(L,t) &=& \frac{1}{2} \left[ |a_1(L_1, t_1)|^2 + |a_2(L_2, t_2)|^2\right] \\
 P^{(2)}_{id}(L_1,t_1;L_2, t_2) &=& \frac{1}{2} \left[  |a_1(L_1, t_1) a_2(L_2, t_2)|^2 + |a_1(L_2, t_2) a_2(L_1, t_1)|^2
 \right. \nonumber \\
 &\pm&
\left.  2 \mbox{Re} \left(a_1(L_1, t_1) a_2^*(L_1, t_1) a_2(L_2, t_2) a_1^*(L_2, t_2)\right) \right],
 \end{eqnarray}

The coincidence function $c^{(2)}(L, t)$ vanishes identically for fermions.   Eq. (\ref{superp2}) is thus violated: fermions exhibit anti-bunching behavior at all times.

For bosons,
 \begin{eqnarray}
 c^{(2)}(L, t) =  \frac{8 |a_1(L, t) a_2(L, t)|^2  }{[|a_1(L,t)|^2+|a_2(L,t)|^2]^2} \label{c2bos}
 \end{eqnarray}
 Plots of $c^{(2)}(L, t)$ are given in Fig. 1 for a specific choice of initial states. Bosons exhibit either bunching or anti-bunching behavior at different times. Fig. 1 also demonstrates that the coincidence function contains information   that is not accessible  from single time-of-arrival measurements.

We consider the special case of a probability current operator $J(L,t)$ of the form (\ref{cur1}). We can always choose the positive operators $\hat{\Pi}_L(t)$ so that the single time probabilities $P^{(1)}(L, t)$ are the same in the POVM and in the probability current description. For example, we can choose $\hat{\Pi}_L(t)$ to be Kijowski' s POVM so that the current operator is a operator-ordered variation of the Schr\"odinger current operator. 
Thus, the only difference lies in the value of the coincidence function. We obtain
\begin{eqnarray}
c^{(2)}_J(L, t) = \frac{1}{2} c^{(2)}_{\Pi}(L, t)  + \frac{\langle \psi_1|\hat{\Pi}_L(t)^2|\psi_1\rangle + \langle \psi_2|\hat{\Pi}_L(t)^2|\psi_2\rangle }{\langle \psi_1|\hat{\Pi}_L(t)|\psi_1\rangle^2 +\langle \psi_1|\hat{\Pi}_L(t)|\psi_1\rangle^2}, \label{c2j}
\end{eqnarray}
where $c^{(2)}_J(L, t)$ is the coherence function for the probability current operator and $c^{(2)}_{\Pi}(L, t)$ is the coherence function evaluated in terms of the POVM as above. The key observation is that the second term in the r.h.s. of Eq. (\ref{c2j}) is always larger than unity since $\langle \hat{\Pi}_L(t)^2\rangle \geq \langle \hat{\Pi}_L(t)\rangle^2$. Since $c^{(2)}_{\Pi}(L, t) \geq 0$, the second term guarantees that that $c^{(2)}_J(L, t)$ is always larger than unity. We note that the term $\langle \hat{\Pi}_L(t)^2\rangle$ diverges if $\hat{\Pi}_L(t)$ is Kijowski's POVM,  the  current operator requires appropriate smearing in order to be well-defined.

The C-S constraint (\ref{bunching}) is always violated for fermions, since $c^{(2)}(L, t) = 0$ even when the coherence function $C^{(2)}(L_1, t_1; L_2, t_2)$ is non zero. In Fig.2, we plot  $C^{(2)}(L_1, t_1; L_2, t_2)$ for a specific choice of initial state. The coherence function is characterized by oscillations, with a characteristic frequency of the order of  $|\bar{E}_1- \bar{E}_2|$, where $\bar{E}_i$ is the mean energy of the state $|\psi_i\rangle$. Whether such oscillations are observable or not depends on the scale $\sigma$ of temporal coarse-graining. The probability density (\ref{pltltid}) is a special case of the probability density (\ref{pltlt2}). This implies that  smearing at a scale of $\sigma$ is required  in order to obtain the  observable probabilities (\ref{pltobs}. Hence, the oscillations of $C^{(2)}(L_1, t_1; L_2, t_2)$ are observable only if $|\bar{E}_1- \bar{E}_2| \sigma$ is at most of order unity.

For bosons, we evaluate the ratio
\begin{eqnarray}
h(t_1, t_2) = \frac{\sqrt{c^{(2)}(L,t_1)c^{(2)}(L,t_2)}}{C^{(2)}(L_1, t_1; L_2, t_2)}, \label{ratioh}
\end{eqnarray}
 which is larger than unity when the C-S constraint is satisfied. In Fig. 3, $h(t_1, t_2)$ is evaluated for a specific initial state. It takes values both larger and smaller than unity, so  the C-S constraint is  violated also for bosons.

 \begin{figure}[tbp]
\includegraphics[height=10cm]{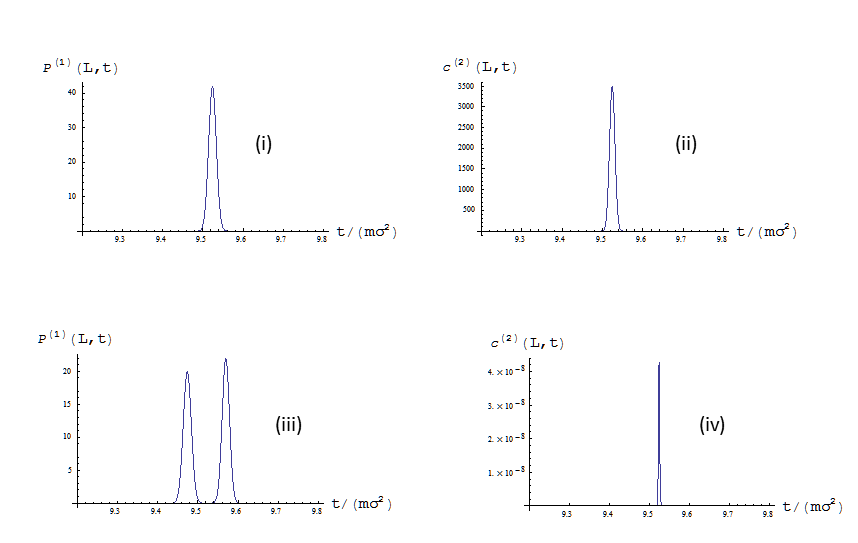} \caption{ \small   Single-time probability densities and coincidence functions for an initial state of the type (\ref{psi02}), where $\psi_i(x) = \phi(x- x_i) e^{i p_i x}$, for some constants $x_i, p_i, i = 1, 2$. We choose for $\phi(x)$ a Gaussian $\phi(x) = (2\pi\sigma_x^2)^{-1/4}exp[-x^2/(4\sigma_X^2)]$. For $|x_2-x_1|>>\sigma_X$, $\psi_1$ and $\psi_2$ are orthogonal. The mean time of arrival for each wave-packet is $\bar{t}_i = m(L-x_i)/p_i$.   We have chosen $L/\sigma_X = 1000, p_1\sigma_X =100, p_2 \sigma_X = 110$.
 In plots (i) and (ii),  $\bar{t}_1 = \bar{t}_2$. The  superposition cannot be identified at the level of the probability density  $P^{(1)}_{id}(L,t) $ of Plot (i). Plot (ii) describes the coincidence function $c^{(2)}(L,t) $ as a function of $t/(m\sigma_X^2)$ for bosons. Bunching or anti-bunching behavior is time-dependent.
Plots (iii) and (iv) are the same as (i) and (ii) only with   $\bar{t}_1 = 0.99 \bar{t}_2$. There are two distinguishable peaks in the probability density, and the peak in  $c^{(2)}(L,t) $ is much lower and narrower.
}
\end{figure}

\begin{figure}[tbp]
\includegraphics[height=5cm]{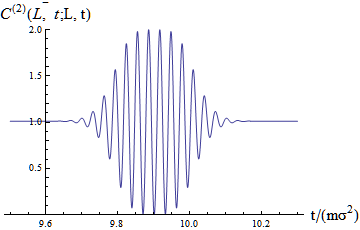} \caption{ \small   Coherence function $C^{(2)}(L, t_1; L, t_2)$ for an initial fermionic state of the type (\ref{psi02}), where $\psi_i(x) = \phi(x- x_i) e^{- p_i x}$, for some constants $x_i, p_i, i = 1, 2$;  $\phi(x)$ is the same Gaussian as in Fig. 1.   We have chosen $L/\sigma_X = 1000, p_1\sigma_X =100, p_2 \sigma_X = 102$, and $x_i$ so that  the mean times of arrival of each wave-packet  have the same value $\bar{t}$. In the plot, we fix $t_1 = \bar{t}$, and vary $t_2$ using dimensionless units.
}
\end{figure}
\begin{figure}[tbp]
\includegraphics[height=5cm]{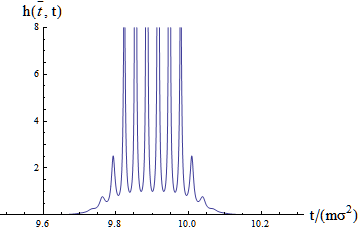} \caption{ \small   The ratio $h(t_1, t_2)$,Eq. (\ref{ratioh}), evaluated for a bosonic initial state of the form   (\ref{psi02}), for the same $\psi_i(x)$ with Fig. 2.   Both wave packets $\psi_i(x)$ have the same mean time of arrival $\bar{t}$. In the plot,  we fix $t_1 = \bar{t}$, and vary $t_2$  using dimensionless units.
}
\end{figure}

\section{Sequential time-of-arrival measurements}
 In this section, we consider two successive time-of-arrival measurements on the same particle. We find how  the quantum state changes after a time-of-arrival measurement and we construct the probabilities for the time-of-flight velocity.
\subsection{Probability assignment}
Two successive  measurements can be performed on a single particle only if the first measurement does not annihilate the particle. In  the QTP method, this implies that the composite operator $\hat{Y}_1({\pmb x})$ of Eq.
  (\ref{pltlt})   must describe particle scattering rather than absorption, i.e.,  $\hat{Y}_1({\pmb x}) = \hat{\psi}^{\dagger}({\pmb x}) \hat{\psi}({\pmb x})$. There is no constraint for the second measurement, so we take $\hat{Y}_2({\pmb x}) = \hat{\psi}({\pmb x})$, as in Secs. 3 and 4.

Eq. (\ref{pltlt}) for an one-particle initial state $|\Psi_0\rangle = \int d^3k \psi_0({\pmb k}) \hat{a}^{\dagger}_{\pmb k}|0\rangle$ yields
 \begin{eqnarray}
 P^{(2)}_{f.g}({\pmb L}_1, t_1; {\pmb L}_2, t_2) = C \int \frac{d^3k_1 d^3 k_2 d^3 k_1' d^3k_2'}{(2\pi)^9}  \; \tilde{g}_1\left(\frac{\epsilon_{{\pmb k}_1} - \epsilon_{{\pmb k}_2} +\epsilon_{{\pmb k}_1'} -\epsilon_{{\pmb k}_2'}}{2}  \right)\tilde{g}_2\left(\frac{\epsilon_{{\pmb k}_2} +\epsilon_{{\pmb k}_2'}  }{2}
 \right)
 \nonumber \\
 \times e^{i({\pmb k}_2 - {\pmb k}_2')\cdot{\pmb L}_2 - i (\epsilon_{{\pmb k}_2} - \epsilon_{{\pmb k}_2}') t_2} e^{i({\pmb k}_1 - {\pmb k}_2 - {\pmb k}_1' + {\pmb k}_2')\cdot{\pmb L}_1 - i (\epsilon_{{\pmb k}_1} - \epsilon_{{\pmb k}_2} -\epsilon_{{\pmb k}_1'} +\epsilon_{{\pmb k}_2'})t_1  } \tilde{\psi}_0({\pmb k}_1') \tilde{\psi}_0({\pmb k}_1).  \label{seqtoa1}
 \end{eqnarray}
 where $\tilde{g}_i$ are the Fourier transforms of the functions $g_i$.

In general, the particle can scatter towards any direction after the first measurement. We restrict to propagation along the axis connecting the source to the locus ${\pmb L}_1$ of the first detector, and then along the axis ${\pmb L}_2 - {\pmb L}_1$ connecting the loci of the two apparatuses. The assumption  that ${\pmb L}_1$ and ${\pmb L}_2$ are parallel incurs no loss of generality and allows us to use a notation for an one-dimensional problem.

For an initial state with positive momentum,  Eq. (\ref{seqtoa1}) becomes
\begin{eqnarray}
 P^{(2)}_{f.g}(L_1, t_1;  L_2, t_2) = C \int \frac{dk_1 d k_2 d k_1' dk_2'}{(2\pi)^3}  \; \tilde{g}_1\left(\frac{\epsilon_{k_1} - \epsilon_{k_2} +\epsilon_{k_1'} -\epsilon_{k_2'}}{2}  \right)\tilde{g}_2\left(\frac{\epsilon_{k_2} +\epsilon_{k_2'}  }{2}
 \right)
 \nonumber \\
\times  \theta(k_2) \theta(k_2') e^{i( k_2 - k_2')\cdot L_2 - i (\epsilon_{k_2} - \epsilon_{k_2'}) t_2} \; e^{i( k_1 - k_2 - k_1' + k_2') L_1 - i (\epsilon_{k_1} - \epsilon_{k_2} -\epsilon_{k_1'} +\epsilon_{k_2'})t_1  } \tilde{\psi}_0^*(k_1') \tilde{\psi}_0(k_1).  \label{seqtoa2}
 \end{eqnarray}
 We have restricted the integration to positive values of $k_2$ and $k_2'$, since particles exiting the first detector with negative momenta with negative  momenta will not be recorded by the second detector\footnote{The positivity of $k_2$ and $k_2'$ is not an additional assumption. We phrased it as such because in this paper we ignore position coarse-graining---we employ the approximation (\ref{poscg}) in the Appendix A.2. Suppose we smear Eq. (\ref{seqtoa2}) with respect to the position $L_1$. If the smearing length is sufficiently large,  a delta function  $\delta(k_1 - k_2 - k_1' + k_2')$ appears. Together with the constraints to the energies from   $\tilde{g}_1$,  it guarantees the positivity of $k_2$ and $k_2'$. The particle does not back-scatter as a result of the first time-of-arrival measurement.
  }.

 The probability densities (\ref{seqtoa2}) are strongly suppressed,   if $t_1 < 0$ or if $t_2 < t_1$. Hence, the total probability $\mbox{Prob}(L_1, L_2)$ that two detection events have occurred is well approximated by
 integrating $P^{(2)}_{f.g}$ along the whole real axis for both $t_1$ and $t_2$. Then,
 \begin{eqnarray}
 \mbox{Prob}(L_1, L_2) := \int_{-\infty}^{\infty} dt_1 \int_{-\infty}^{\infty}dt_2  P^{(2)}_{f.g}(L_1, t_1;  L_2, t_2)
 \nonumber \\
 = Cm^2  \int \frac{dk_1 dk_2}{2\pi|k_1 k_2|} \tilde{g}_1(\epsilon_{k_1} - \epsilon_{k_2}) \tilde{g}_2(\epsilon_{k_2}) \theta(k_2) |\tilde{\psi}_0(k_1)|^2. \label{prl1l2}
 \end{eqnarray}

 Eq. (\ref{prl1l2}) implies that
 \begin{enumerate}[(i)]
\item  $\tilde{g}_2(\epsilon_k)/|k|$ is the absorption coefficient of the second detector,  and
\item $\frac{1}{|k_1|}\tilde{g}_1(\epsilon_{k_1} - \epsilon_{k_2})$ is the probability that an incoming particle of momentum $k_1$ is scattered to a different momentum $k_2$.
 \end{enumerate}

 We consider ideal detectors. For the first detector, the idealization consists in the assumption that energy transfer during scattering is negligible. For the second detector, the idealization is the same with Sec. 3.1., i.e., we assume that   particle absorption in the second detector is independent of the particle's momentum. These conditions imply that $\tilde{g}_1(\epsilon_{k}-\epsilon_{k'}) \sim \epsilon_k \delta (\epsilon_k - \epsilon_{k'})$ and that $g_{2}(\epsilon) \sim \sqrt{\epsilon}$. Choosing the constant $C$ so that  $\mbox{Prob}(L_1, L_2) = 1$, we obtain the ideal probability distribution
 \begin{eqnarray}
  P^{(2)}_{id}(L_1, t_1;  L_2, t_2) = \int \frac{dk_1 d k_2 d k_1' dk_2'}{2\pi^2 } \left(\frac{\epsilon_{k_1}+\epsilon_{k_1'}}{m}\right)^{3/2} \delta( \epsilon_{k_1} - \epsilon_{k_2} +\epsilon_{k_1'} -\epsilon_{k_2'})   \theta(k_2) \theta(k_2')\nonumber \\
  \times e^{i( k_2 - k_2')\cdot L_2 - i (\epsilon_{k_2} - \epsilon_{k_2'}) t_2} \; e^{i( k_1 - k_2 - k_1' + k_2') L_1 - i (\epsilon_{k_1} - \epsilon_{k_2} -\epsilon_{k_1'} +\epsilon_{k_2'})t_1  } \tilde{\psi}_0^*(k_1') \tilde{\psi}_0(k_1). \label{seqideal}
 \end{eqnarray}

Eq.  (\ref{seqideal})  does not have the standard form of probability densities for sequential measurements. If  the observable $A$ corresponding to a  POVM $\hat{E}(a)$ is measured first, and the observable $B$ corresponding to a POVM $\hat{F}(b)$ is measured second, the joint probability density is
\begin{eqnarray}
P(a, b) = Tr \left(\hat{\rho}_0 \sqrt{\hat{E}}(a) \hat{F}(b)\sqrt{\hat{E}}(a)\right), \label{pabe}
\end{eqnarray}
where $\hat{\rho}_0$ is the initial state.   Eq. (\ref{seqideal}) cannot be brought in the form (\ref{pabe}). This is not surprising since the two measurements in Eq. (\ref{pabe}) take place at fixed time instants, while time is a random variable in time-of-arrival measurements.

\subsection{Marginal distributions}
The two marginal distributions of the probability density(\ref{seqideal}) have different properties. When tracing out the time $t_2$ of the second measurement, we recover the probability distribution $P^{(1)}_{id}(L_1, t_1)$ of Eq. (\ref{pltid}),
\begin{eqnarray}
\int_{-\infty}^{\infty} dt_2   P^{(2)}_{id}(L_1, t_1;  L_2, t_2)  = P^{(1)}_{id}(L_1, t_1).
\end{eqnarray}
By causality,  the second measurement cannot affect the statistics of the first one.

However, when tracing out the time $t_1$ of the first measurement, we obtain a probability distribution that differs from $P^{(1)}_{id}(L_2, t_2)$ ,
\begin{eqnarray}
\int_{-\infty}^{\infty} dt_1   P^{(2)}_{id}(L_1, t_1;  L_2, t_2) =  \int \frac{dkdk'}{2\pi } \frac{\sqrt{m}(\epsilon_k+\epsilon_{k'})^{3/2}}{|kk'|} e^{i( k - k')L_2 - i (\epsilon_{k} - \epsilon_{ k'})t_2}\tilde{\psi}_0^*(k') \tilde{\psi}_0(k). \label{marg1}
\end{eqnarray}
The marginal distribution (\ref{marg1}) is of the form (\ref{pltidd}) for
\begin{eqnarray}
\eta(x) = \frac{(1+\frac{x^2}{16})^{3/2}}{1-\frac{x^2}{16}}.
\end{eqnarray}
Eq. (\ref{marg1}) implies that the  first measurement has transformed the initial state $\hat{\rho}_0$ as
\begin{eqnarray}
\hat{\rho}_0 \rightarrow   \left( \hat{p}\hat{\rho}_0\hat{p}^{-1} + \hat{p}^{-1}\hat{\rho}_0\hat{p}\right).
\end{eqnarray}

In order to find the analogue of the state reduction for a time-of-arrival measurement, we write
  Eq. (\ref{seqideal})   as
\begin{eqnarray}
P^{(2)}_{id}(L_1, t_1;  L_2, t_2) = \int \frac{dkdk'}{2\pi } \sqrt{\frac{\epsilon_k+\epsilon_{k'}}{m}} e^{i( k - k')(L_2 - L_1) - i (\epsilon_{k} - \epsilon_{ k'})(t_2 - t_1)} \rho^{(red)}_{L_1, t_1}(k,k'),
\end{eqnarray}
where
\begin{eqnarray}
 \rho^{(red)}_{L, t}(k,k') = \frac{\epsilon_k + \epsilon_k'}{m} \int \frac{dk_1dk'_1}{\pi} \delta(\epsilon_k+\epsilon_{k'} - \epsilon_{k_1}-\epsilon_{k_1'}) e^{i(k_1-k_1')L - i (\epsilon_{k_1}-\epsilon_{k_1'}) t} \rho_0(k, k')
 \nonumber \\
= \frac{\epsilon_k +\epsilon_{k'}}{\pi m}  \int_{-\infty}^{\infty}ds e^{is(\epsilon_k +\epsilon_{k'})} \langle L |e^{i\hat{H}(t+s)} \hat{\rho}_0e^{-i\hat{H}(t-s)}|L\rangle
\end{eqnarray}
Thus, a  measurement by a detector at $L$ that records the value $t$ changes the state of the system, by a generalized `state reduction' rule
\begin{eqnarray}
\hat{\rho}_0 \rightarrow \hat{\rho}^{(red)}_{L, t} = \frac{1}{\pi m}  \int_{-\infty}^{\infty}ds\langle L |e^{i\hat{H}(t+s)} \hat{\rho}_0e^{-i\hat{H}(t-s)}|L\rangle   \left[\hat{H}|s\rangle\langle - s| + |s\rangle\langle -s|\hat{H}\right]  , \label{redux}
\end{eqnarray}
where $|s\rangle = \int_0^{\infty} dk e^{is \epsilon_k}|k\rangle$.

Obviously, the rule (\ref{redux}) is very different from the standard rule of quantum state reduction.    This is not surprising,  because a time-of-arrival measurement  refers to a fixed point in space and variable time, in contrast to the usual reduction rule that refers to a fixed instant of time. We note that the transformation (\ref{redux})  is constructed solely from the Hamiltonian $\hat{H}$ and the generalized eigenstates of the position operator.

\subsection{Classical correspondence}
 We rewrite Eq. (\ref{seqideal})  as
\begin{eqnarray}
  P^{(2)}_{id}(L_1, t_1;  L_2, t_2) = \int dk dk' \psi_0^*(k') \psi_0(k) \langle k|\hat{\Pi}_{L_1}(t_1)|k' \rangle F(\epsilon_{k}+\epsilon_{k'}, L_2-L_1, t_2-t_1),  \label{seqid2}
\end{eqnarray}
where $\langle k|\hat{\Pi}_{L_1}(t)|k' \rangle $ is given by Eq. (\ref{pilt}) and
\begin{eqnarray}
F(E, \ell, \tau) = \frac{2E}{\pi m} \int_0^{\infty} dk \int_0^{\infty} dk' e^{i(k-k')
\ell - i (\epsilon_k - \epsilon_{k'})\tau} \delta (\epsilon_k+\epsilon_{k'}-E). \label{Felt}
\end{eqnarray}
The Fourier transform of the function $F$ with respect to $\tau$ is readily evaluated,
\begin{eqnarray}
\tilde{F}(E, \ell, \mu) := \left\{\begin{array}{cc} \int d\tau e^{-i \mu \tau} F(\epsilon, \ell, \tau) = \frac{ E}{\sqrt{E^2 - \mu^2}} e^{i\sqrt{m(E - \mu)} \ell - \sqrt{m(E + \mu)}\ell}, & \mu \leq E\\0& \mu > E \end{array}\right..
\label{Felt2}
\end{eqnarray}

 In the Appendix B.2, we obtain  an analytic expression for the function $F$, using a stationary phase approximation to the inverse Fourier transform of $\tilde{F}$.   Here, we note that for $|\mu| << E$, $\tilde{F}(E, \ell, \mu) \simeq e^{-i \sqrt{m/E}\ell \mu}$. Hence, for sufficiently large times ($E \tau >> 1$), $F$  approximates  a delta function,
\begin{eqnarray}
F(E, \ell, \tau) \simeq \delta (\tau - \sqrt{\frac{m}{E}}\ell).  \label{fcl}
\end{eqnarray}
Transforming Eq. (\ref{seqid2}) into the Wigner picture, we obtain
\begin{eqnarray}
  P^{(2)}_{id}(L_1, t_1;  L_2, t_2) =  \int dX dP W_0(X, P) \nonumber \\ \times \left(\frac{1}{2\pi m} \int d \xi e^{i \xi (L_1 - X-\frac{P}{m}t_1)} \sqrt{P^2+\frac{\xi^2}{4}} F(\frac{P^2}{m} + \frac{\xi^2}{4m}, L_2-L_1, t_2-t_1) \right). \label{pseqw}
\end{eqnarray}

An inspection of Eq. (\ref{pwid}) shows that the classical time-of-arrival probability distribution corresponds to the limit $|\xi|<<|P|$ in the integral. This is the same regime in which Eq. (\ref{fcl}) applies.  Approximating $P^2+\xi^2/4 \simeq P^2$ and using Eq. (\ref{fcl}),  we obtain the classical probability distribution for two successive time-of-arrival measurements
\begin{eqnarray}
  P^{(2)}_{cl}(L_1, t_1;  L_2, t_2) =  \int dX dP \; W_0(X, P)  \delta(t_1- T_{c1})  \delta (t_2- t_1 - T_{c2} +T_{c1}), \label{pseqwcl}
\end{eqnarray}
where
\begin{eqnarray}
T_{ci} = m\frac{L_i - X}{P}.
\end{eqnarray}

\subsection{Time-of-flight velocity}

Next, we define the probability density $P(\tau)$ for the {\em time-of-flight} $\tau = t_2 - t_1$ between the two measurements
\begin{eqnarray}
P(\tau) :=   \int dt P^{(2)}_{id}(L_1, t;  L_2, t + \tau) = \int dk |\tilde{\psi}_0(k)|^2 F(2\epsilon_k, L_2 - L_1, \tau).  \label{toft}
\end{eqnarray}
We use the term `time of flight' as distinct from the term `time of arrival'. The time of arrival refers to one measurement record on a single detector, while the time of flight requires two measurement records at spatially separated detectors.

 The time of flight $\tau$ and the  time $t_1$ of the first measurement are uncorrelated: the correlation function $C_{t_1\tau}$, calculated from Eq. (\ref{seqid2}), vanishes.

Eq. (\ref{toft}) implies that the time of flight can be represented by an operator $\hat{\tau}$ that is a function of the momentum $\hat{p}$: $\hat{\tau} = \tau_f(|\hat{p}|)$, where
\begin{eqnarray}
\tau_{f}(p) = \int_0^{\infty} ds s F(\frac{p^2}{m}, L_2 - L_1, s). \label{tof}
\end{eqnarray}

We also define  the {\em time-of-flight velocity}
\begin{eqnarray}
v_{tof} (p):= \frac{L_2 - L_1}{\tau_f(p)}. \label{vtof}
\end{eqnarray}
%By Eq. (\ref{toft}, the time-of-flight velocity can be expressed as a function of the momentum  operator $v_{tof}(\hat{p})$, where
%\begin{eqnarray}
%v_{tof}(p) = \int_0^{\infty} ds \frac{L_2-L_1}{s}  F(\frac{p^2}{m}, L_2 - L_1, s). \label{vtof}
%\end{eqnarray}

In the regime where   Eq. (\ref{fcl}) applies, Eq. (\ref{tof}) yields $\tau_f(p) = m(L_2-L_1)/p$. Hence,
\begin{eqnarray}
v_{tof}(p) = \frac{p}{m}, \label{vtof2}
\end{eqnarray}
i.e., the time-of-flight velocity coincides with the canonical velocity $p/m$. This result agrees with the classic analysis of Park and Margenau \cite{PaMa68}, even though the context is slightly different. The difference is that the time-of-flight velocity is defined here in terms of two measurements of time at specific locations, while in Park and Margenau's work, the time-of-flight velocity is defined in terms of two position measurements at pre-specified times.

In general, probabilities associated to velocity differ from the probabilities associated to momentum. This is because momentum and velocity are defined differently in any histories-based theory   \cite{Sav99}. This difference is manifested in temporally extended measurements. For example, in  time-extended von Neumann measurements, velocity and momentum correspond to different Hamiltonian operators for the interaction between the quantum system and the measurement apparatus \cite{AnSav07}.

This difference turns out to be insignificant for  time-of-flight velocities. The evaluation of (\ref{vtof}) using the approximation (\ref{Feltairy}) of the Appendix B.2 yields appreciable differences from Eq. (\ref{vtof2}) only for $(L_2 - L_1) p << 1$, i.e., when the distance between the detectors is {\em much smaller} than the de Broglie wavelength of the particles.

\section{Conclusions}
The main result of this paper is the construction of time-of-arrival probabilities for multi-partite systems and for sequential measurements. This was made possible by the use of the QTP method, in which the relevant probabilities are constructed as linear functionals of appropriate field correlation functions. The method can be straightforwardly generalized to set-ups more elaborate than the ones considered here, involving three or more detection events.

When the time-of-arrival problem is formulated in terms of a single particle and a single detector, all approaches lead to almost indistinguishable predictions. In multi-partite systems, this is no longer the case. Approaches to the time of arrival based on probability currents lead to quantitatively different predictions from those based on POVMs.  Therefore, we propose that measurements of  time-of-arrival correlations in multi-partite systems can  distinguish between different theories about the time of arrival. In particular, the QTP method makes a specific prediction about the time-of-arrival  POVM in multi-partite systems.

Besides the main result above, we also showed that the time-of-arrival correlations can play the role of entanglement witness in multi-partite systems. This result suggests that temporal observables can be used for the retrieval of quantum information and might even be employed in quantum information processing. We also  derived the probability distribution associated to sequential time-of-arrival measurements, and we found that the state-reduction rule for a time-of-arrival measurement is very different from the standard one.

\newpage

\begin{appendix}
\section{The Quantum Temporal Probabilities method}
\subsection{General probability assignment}
We consider a composite physical system that consists of a microscopic and a macroscopic component. The microscopic component is the quantum system to be measured and the macroscopic component is the measuring device.

We denote the Hilbert space associated to the composite system
 by
   ${\cal H}$. We describe a measurement event as a transition between two complementary subspaces of ${\cal H}$. To this end, we split
 ${\cal H}$  in two subspaces: ${\cal H} = {\cal
H}_+ \oplus {\cal H}_-$. The subspace ${\cal H}_+$ describes the accessible states of the system given that the event under consideration is realized. For example, if the event is a detection of a microscopic particle by  an  apparatus,   ${\cal H}_+$ corresponds to all states of the apparatus compatible with the macroscopic record of detection.  We
denote  the projection operator onto ${\cal H}_+$ as $\hat{P}$ and the projector onto ${\cal H}_-$ as $\hat{Q} := 1  - \hat{P}$. We assume that the system is described by a Hamiltonian operator $\hat{H}$.

In Refs. \cite{AnSav12, AnSav15}, we constructed the
 probability density with respect to time that is associated to the transition of the system from ${\cal H}_-$ to ${\cal H}_+$.
 A pointer variable $\lambda$ of the measurement apparatus was also assumed to take a definite value after the transition has occurred. It is described by a set of positive operators
$\hat{\Pi}(\lambda)$  that correspond to the different values of $\lambda$.  The operators $\hat{\Pi}(\lambda)$ satisfy
 $  \sum_\lambda \hat{\Pi}(\lambda) = \hat{P}$.

 First, we construct
 the probability amplitude $| \psi; \lambda, [t_1, t_2] \rangle$ that, given an initial ($t=0$) state $|\psi_0\rangle \in {\cal H}_-$, a transition occurs during the time interval $[t_1, t_2]$ and a value $\lambda$ for the pointer variable is obtained
 for
 some observable. For a vanishingly small time
interval, i.e., $t_1 = t$ and $t_2 = t + \delta t$,  one obtains \cite{AnSav12}

\begin{eqnarray}
|\psi_0; \lambda, [t, t+ \delta t] \rangle =  - i \, \delta t \, \,e^{-i\hat{H}(T - t)} \sqrt{\hat{\Pi}}(\lambda) \hat{H} \hat{S}_t |\psi_0
\rangle,  \label{amp1}
\end{eqnarray}
where $
 \hat{S}_t =  \lim_{N
\rightarrow \infty} (\hat{Q}e^{-i\hat{H} t/N} \hat{Q})^N$ is the restricted propagator in the subspace ${\cal H}_-$.

 The amplitude (\ref{amp1}) defines  a {\em density} with respect to time:  $ |\psi_0;  \lambda, t \rangle = - i
e^{- i \hat{H} T} \hat{C}(\lambda, t) |\psi_0 \rangle$, where   $ \hat{C}(\lambda, t) := e^{i \hat{H}t} \sqrt{\hat{\Pi}}(\lambda)\hat{H}\hat{S}_t$ is a {\em history operator}.
The total amplitude that the transition occurred at {\em some moment} within a time interval $[t_1, t_2]$ is

\begin{eqnarray}
| \psi; \lambda, [t_1, t_2] \rangle = - i e^{- i \hat{H}T} \int_{t_1}^{t_2} d t \hat{C}(\lambda, t) |\psi_0 \rangle. \label{ampl5}
\end{eqnarray}

Eq. (\ref{ampl5}) involves the restricted propagator $\hat{S}_t$  which may be difficult to evaluate in practice. We sidestep the evaluation of $\hat{S}_t$, by using the following approximation. We consider a Hamiltonian   $\hat{H} = \hat{H_0} + \hat{H_I}$ where
$[\hat{H}_0, \hat{P}] = 0$, and $H_I$ is a perturbing interaction. To leading order in the interaction,
\begin{eqnarray}
 \hat{C}(\lambda, t) = e^{i \hat{H}_0t} \sqrt{\hat{\Pi}}(\lambda) \hat{H}_I e^{-i \hat{H}_0t}, \label{classpert}
\label{perturbed}
\end{eqnarray}
with no dependence on $\hat{S}_t$. All models for relativistic measurements we consider in this paper use the approximation (\ref{perturbed}).

We construct a probability measure from the amplitude (\ref{ampl5}) by coarse-graining the time variable \cite{AnSav12, AnSav15}. This is a natural procedure  for  systems that involves a macroscopic component such as a measuring apparatus  \cite{GeHa, hartlelo}. Hence, probabilities are defined only for time intervals $]t_1, t_2]$ such that $|t_2 - t_1| >> \sigma$,  where $\sigma$ is the coarse-graining scale.

  We implement temporal coarse-graining  by  defining smeared history operators, $\hat{C}_f(\lambda ,t) = \int ds \sqrt{f}(t-s) \hat{C}(\lambda, s)$. The function $f(s)$ is positive, it is centered around $s = 0$ and has  width of order $\sigma$, like, for example, the Gaussian (\ref{gauss}). The probability density  that a transition occurs during the time interval $[t_1, t_2]$ and a value $\lambda$ for the pointer variable is obtained is
\begin{eqnarray}
P(\lambda, t) = Tr [\hat{C}_f(\lambda, t) \hat{\rho}_0\hat{C}_f^{\dagger}(\lambda, t)],
\end{eqnarray}
where $\hat{\rho}_0 = |\psi_0\rangle \langle \psi_0|$.

 An analogous equation holds for multiple events. The probability density that one event associated to a measurement record $\lambda_1$ occurs at time $t_1$,   and  another  event  associated to a measurement record $\lambda_2$occurs at time $t_2$ is
 \begin{eqnarray}
 P(\lambda_1, t_1;\lambda_2, t_2) = Tr \left[ \hat{C}_{f_1,f_2} (\lambda_1, t_1, \lambda_2, t_2) \hat{\rho}_0 \hat{C}_{f_1,f_2} (\lambda_1, t_1, \lambda_2, t_2)\right], \label{multiplet}
 \end{eqnarray}
where
\begin{eqnarray}
 \hat{C}_{f_1,f_2} (\lambda_1, t_1, \lambda_2, t_2) = \int ds_1 ds_2 f_1(t_1-s_1) f_2(t_2-s_2) \hat{C}(\lambda_1, t_1, \lambda_2, t_2)
\end{eqnarray}
is the smeared form of the amplitude operator
\begin{eqnarray}
\hat{C}(\lambda_1, t_1, \lambda_2, t_2) = {\cal T} ([\hat{C}_2(\lambda_2, t_2) \hat{C}_1(\lambda_1, t_1)], \label{2ampl}
\end{eqnarray}
where ${\cal T}$ is the standard  time-ordering operator and $\hat{C}_i(\lambda_i, t_i)$ are the class operators (\ref{classpert}) for a the $i$-th event.

\subsection{Time of arrival probabilities}
 Next, we specialize to time of arrival measurements. In this case, the measured quantum system consists of free particles and the records of observation $\lambda$ are identified with the location ${\pmb L}$ of a particle detector.

Let  ${\cal F}$ be the Hilbert space associated to the particles. For treating  multi-particle states, it is convenient to identify ${\cal F}$ with a Fock space, either bosonic or fermionic.  Hence,  ${\cal F}$ carries either a representation of the canonical commutation relations (\ref{ccr}), or of the canonical anti-commutation relations (\ref{car}). We will denote the Hamiltonian on ${\cal F}$ by $\hat{h}$ and the initial state of the field by $|\Psi_0\rangle$.  Ignoring spin and internal degrees of freedom, and restricting to non-relativistic particles, the  field operators on ${\cal F}$ are defined by Eq. (\ref{fields}).

We assign a Hilbert space ${\cal K}_i$ to each detector. Thus, for a single detection event, the Hilbert space of the total system is ${\cal H}_1 = {\cal F} \otimes {\cal K}_1$ and for two detection events the Hilbert space of the total system is ${\cal H}_2 = {\cal F} \otimes {\cal K}_1\otimes {\cal K}_2$. We will denote the initial state of each detector as $|\Phi_0^{(i)}\rangle$.

We make the common assumption in von Neumann measurements that the self-dynamics of the detector is negligible. This implies that the unperturbed Hamiltonian $\hat{H}_0$ is $\hat{h} \otimes \hat{1}$ on ${\cal H}_1$ and $\hat{H}_0 = \hat{h} \otimes \hat{1}\otimes \hat{1}$ on ${\cal H}_2$. We assume a local  Hamiltonian governing the interaction between the particles and the detector of the form $\hat{H}_I = \int d^3x \hat{Y}_a({\pmb x}) \otimes J^a({\pmb x})$, where $\hat{Y}_a$ are composite operators on ${\cal F}$ and $\hat{J}^a({\pmb x})$ are current operators on the Hilbert space of the detector.

Two cases are of particular interest.
\begin{enumerate}
\item An interaction Hamiltonian linear with respect to the fields
\begin{eqnarray}
\hat{H}_I = \int d^3x \left[ \psi({\pmb x}) \otimes \hat{J}({\pmb x}) + \hat{\psi}^{\dagger}({\pmb x}) \otimes \hat{J}^{\dagger}({\pmb x})\right].
 \end{eqnarray}
 This Hamiltonian corresponds to detection by particle absorption. In order to avoid spurious signals in the detector,  it is necessary to assume that the initial state $|\Phi_0\rangle$ of the apparatus satisfies $\hat{J}^{\dagger}|\Phi_0\rangle = 0$. Hence,  $\hat{Y}$ effectively coincides with $\psi({\pmb x})$.

 \item An  interaction Hamiltonian  $\hat{H}_I = \int d^3x \hat{\psi}^{\dagger}({\pmb x})   \hat{\psi}({\pmb x}) \otimes \hat{J}({\pmb x})$ that is quadratic to the fields corresponds to the scattering of the detected particle. The associated composite operator $\hat{Y}({\pmb x})$ is the particle density $\hat{\psi}^{\dagger}({\pmb x})   \hat{\psi}({\pmb x})$.
\end{enumerate}

The interaction Hamiltonian of the two-detector system is
\begin{eqnarray}
\hat{H}_I = \int d^3 x \left( \hat{Y}_1({\pmb x}) \otimes \hat{J}_1({\pmb x}) \otimes \hat{1} + \hat{Y}_2({\pmb x}) \otimes \hat{1}  \otimes \hat{J}_2({\pmb x})\right).
\end{eqnarray}

The detectors are assumed to be static and localized in position. Thus, a detection event corresponds to a position pointer variable ${\pmb L}$, described by positive operators $\hat{P}_{\pmb L}$ on the Hilbert space ${\cal K}$ of the detector. This implies that the positive operator $\hat{\Pi}_{\lambda}$ in Eq. (\ref{classpert}) is $\hat{1} \otimes \hat{P}_{\pmb L}$ on ${\cal H}_2$.

Assuming an initial state $|\Psi_0\rangle \in {\cal F}$ for the particles and an initial state $|\Phi_0\rangle \in {\cal K}_1$ for the detector, Eq. (\ref{classpert}) yields
\begin{eqnarray}
P({\pmb L}, t) = \int ds ds' \sqrt{f(t-s)f(t-s')} \int d^3x d^3x' \langle \Psi_0| Y^{\dagger}({\pmb x'},s') \hat{Y}({\pmb x}, s) |\Psi_0\rangle \nonumber \\
\times \langle \Phi_0|\hat{J}^{\dagger}({\pmb x'}) \hat{P}_{\pmb L}\hat{J}({\pmb x})|\Phi_0\rangle. \label{plt1b}
\end{eqnarray}

If the distance between the particle source from the detector is much larger than the size of the detector, $\hat{P}_{\pmb L} \hat{J}({\pmb x}) |\Phi_0\rangle$ vanishes unless ${\pmb L} \simeq {\pmb x}$. Ignoring position coarse-graining, we can approximate
\begin{eqnarray}
\langle \Phi_0|\hat{J}^{\dagger}({\pmb x'}) \hat{P}_{\pmb L}\hat{J}({\pmb x})|\Phi_0\rangle = C \delta^3 ({\pmb x} - {\pmb L})\delta^3 ({\pmb x} - {\pmb L}), \label{poscg}
\end{eqnarray}
for some positive constant $C$. Hence, Eq. (\ref{plt1b}) gives Eq. (\ref{plt1}).

Using the  same procedure, one obtains  Eq. (\ref{pltlt}) from Eq. (\ref{multiplet}).

\section{Functions that appear in probability distributions}
\subsection{The distribution  $u(s)$, Eq. (\ref{ws})}
The integral $u(s)$, Eq. (\ref{ws}), defines a probability distribution that is singular at $s = 0$, and it is characterized by a moment-generating functional
\begin{eqnarray}
z(\mu) := \int ds u(s) e^{i \mu s} = \sqrt{1 + \mu^2}
\end{eqnarray}

The moments of $u(s)$ are obtained from the differentiation of $Z(\mu)$ at $\mu = 0 $,
\begin{eqnarray}
\int ds s^n u(s) = \left\{ \begin{array}{cc} \frac{1}{2^k}&n=2k, k=0, 1, 2. \\
\frac{1 \cdot 3 \cdot \ldots \cdot (2k-3)}{2^k}& n= 2k, k = 3, 4, \ldots \\
0& n = 2k+1, k=0,1,2, \ldots. \end{array} \right. \label{intut}
\end{eqnarray}

The distribution $u(s)$ is best characterized as
\begin{eqnarray}
u(s) = \frac{d}{ds} \zeta(s)
\end{eqnarray}
where $\zeta(s)$  is the weak limit of the family of  functions
\begin{eqnarray}
\zeta_{\epsilon}(s)  = \frac{1}{\pi} \int_0^{\infty} dk \frac{\sin ks}{k}\sqrt{1+k^2}e^{-k \epsilon},  \label{zetas}
\end{eqnarray}
as $\epsilon \rightarrow 0$.

For large values of $|s|$, the dominant contribution to the integral  (\ref{zetas}) is
\begin{eqnarray}
\frac{1}{\pi}  \int_0^{\infty} dk \frac{\sin ks}{k} , \nonumber
\end{eqnarray}
  which has a finite value $\frac{1}{2} \mbox{sgn}(s) $. Hence $\zeta(s) \simeq \theta (s) - \frac{1}{2}$, where $\theta(s)$ is the Heavyside step function. This  approximation that justifies the substitution of $u(s)$ with a delta function, since $\delta(s) = \frac{d}{ds} \theta(s)$.

For $s$ around zero,
\begin{eqnarray}
\zeta_{\epsilon}(s) \simeq \frac{1}{\pi} \int_0^{\infty} dk \sin ks e^{-k \epsilon} = \frac{1}{\pi}  \frac{s}{s^2+ \epsilon^2}.  \label{zetas2}
\end{eqnarray}
Eq. (\ref{zetas2}) implies that in the vicinity of $s = 0$,   $\zeta$ approaches the distribution $\mbox{PV}\frac{1}{s}$, where $\mbox{PV}$ stands for the Cauchy principal value.

\begin{figure}[tbp]
\includegraphics[height=6cm]{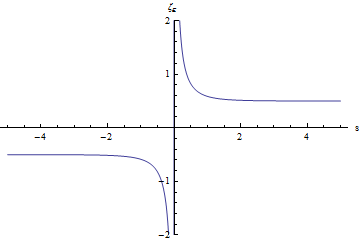} \caption{ \small  Graph of the function $\zeta_{\epsilon}(s)$ of Eq. (\ref{zetas}) for $\epsilon = 10^{-5}$.}
\end{figure}

The distribution $\zeta(s)$ decreases everywhere with $s$ except for the infinite jump from $- \infty$ to $ \infty$ at $s = 0$.  Even if $u(s)< 0$ everywhere but $s = 0$, the positive contribution from the infinite jump dominates over the negative values so that the moments (\ref{intut}) turn out to be positive. However, the integral $\int_{-\infty}^{\infty} ds u(s) f(s)$ is negative for any smooth positive function $f(s)$ that vanishes in an open neighborhood of $s = 0$.

Regarding the role of $u(s)$ in the probability density (\ref{pwid}), we note the following. If Eq. (\ref{pwid}) involved a classical probability distribution rather than a Wigner function, the probabilities Eq.(\ref{pwid}) could become negative. A classical probability distribution can have support  on a bounded region $U$ of the $(X, P)$ plane, so that the argument of $u$ in Eq. (\ref{pwid}) cannot becomes  zero  for $(X, P) \in U$. However, unlike a classical probability distribution, a Wigner function cannot be sharply localized in a bounded region. Since the argument of $u$ in Eq. (\ref{pwid}) involves both $X$ and $P$, it will always cross $s =0$, even if $s = 0$ corresponds to a tail of the Wigner function. Hence, the densities (\ref{pwid}) are positive for Wigner functions, but they can be negative for classical probability distributions that are not Wigner functions of some quantum state.

\subsection{The function $F(E, \ell, \tau)$, Eq. (\ref{Felt})}
We compute $F(E, \ell, \tau)$ by evaluating the inverse Fourier transform of Eq. (\ref{Felt2}). Setting $x = \mu/E$, we find
\begin{eqnarray}
F(E, \ell, \tau) = \frac{E}{2\pi} \int_{-1}^1 \frac{dx}{\sqrt{1-x^2}}e^{i(E\tau)S(x)}, \label{Felt3}
\end{eqnarray}
where
\begin{eqnarray}
S(x) = x - \gamma (\sqrt{1+x}-\sqrt{1-x}),
\end{eqnarray}
for
\begin{eqnarray}
\gamma = \sqrt{\frac{m}{E}} \frac{\ell}{\tau}.
\end{eqnarray}
For $\gamma < 1$, the function $S(x)$ has two critical points at $x = \pm x_0$, where
\begin{eqnarray}
x_0 =  \sqrt{1 - \frac{\gamma^2}{4}(1+\sqrt{1+8/\gamma^2}) }.
\end{eqnarray}
For $\gamma > 1$, there are no critical points.

We evaluate the integral in Eq. (\ref{Felt3}), using the stationary phase method, to obtain
\begin{eqnarray}
F(E, \ell, \tau) =  \sqrt{\frac{2E }{\pi B(\gamma)\tau} }\cos[A(\gamma)E\tau  - \frac{\pi}{4}], \label{spa}
\end{eqnarray}
for $\gamma \leq 1$.
We used the notation $A(\gamma)  =  S(x_0)$ and $B(\gamma) = |S''(x_0)|(1-x_0^2)$.
For $\gamma > 1$, $F$ is strongly suppressed. Within the domain of validity if Eq. (\ref{spa}), we can set $F(E, \ell, \tau) = 0$.

For $\gamma$ close to $1$, $x_0 \simeq \sqrt{\frac{4}{3}(1-\gamma)}$, and $B(\gamma) \simeq  \sqrt{\frac{3}{4}(1-\gamma)}$. Hence, $F(E, \ell, \tau)$ is sharply peaked at $\gamma = 1$. When evaluating integrals of the form  $\int_0^{\infty} d\tau F(E, \ell, \tau) g(\tau)$
for some positive function $\tau$, only values of $\gamma$ close to unity contribute significantly. In this case, the critical points of $S(x)$ are close to zero, so we are justified in expanding $S(x)$ around $x = 0$. The lowest order in the expansion leads to the classical limit,  Eq. (\ref{fcl}). Keeping terms up to the next order, we approximate
\begin{eqnarray}
F(E, \ell, \tau) \simeq \frac{E}{2\pi} \int_{-\infty}^{\infty} dx e^{iE\tau[x - \gamma( x + \frac{1}{8}x^3)]} = D \; \mbox{Ai}[-D(\tau - \sqrt{m/E}\ell)], \label{Feltairy}
\end{eqnarray}
where $\mbox{Ai}$ is the Airy function and $D = 2\left(\frac{E^5}{9m\ell^2}\right)^{1/6}$.

\end{appendix}

 \end{document}